\documentclass[twocolumn, times]{aastex631}
\usepackage{graphicx}   
\usepackage{amsmath}	
\usepackage{hyperref}   
\usepackage[T1]{fontenc} 
\usepackage{mathptmx}
\usepackage{txfonts}
\usepackage{xspace}
\usepackage{gensymb} 
\usepackage{numprint}
\usepackage[normalem]{ulem}

\npthousandsep{\,}
\def\la{{\rm Lyman-}$\alpha$\xspace}
\def\hi{\textsc{Hi}\xspace}
\def\hiim{\textsc{Hi\xspace IM}\xspace}

\def\ob{\Omega_{\rm B}}

\def\nuhi{\nu_{\rm \hi}}
\def\vpec{v_{\rm pec}}
\def\aeff{A_{\rm eff}}
\def\tsys{T_{\rm sys}}
\def\tsky{T_{\rm sky}}
\def\trec{T_{\rm rec}}

\def\mhz{\rm MHz}
\def\ghz{\rm GHz}

\def\hmpc{h^{-1}{\rm Mpc}}
\def\kpc{\rm kpc}

\def\mk{\rm mK}

\def\mgspairs{76266 MGS pairs }

\newcommand{\add}[1]{{#1}}

\newcommand{\reffg}[1]{Figure~\ref{#1}}
\newcommand{\reftb}[1]{Table~\ref{#1}}
\newcommand{\refeq}[1]{Equation~(\ref{#1})}
\newcommand{\refsc}[1]{Section~\ref{#1}}
\newcommand{\refap}[1]{Appendix~\ref{#1}}

\begin{document}
\title{FAST drift scan survey for \hi intensity mapping: simulation on hunting \hi filament with pairwise stacking}

\author[0009-0000-6895-9136]{Diyang Liu}
\affiliation{Key Laboratory of Cosmology and Astrophysics (Liaoning) \& College of Sciences, Northeastern University, Shenyang 110819, China}

\author[0000-0003-1962-2013]{Yichao Li}
\correspondingauthor{Yichao Li}
\email{liyichao@mail.neu.edu.cn}
\affiliation{Key Laboratory of Cosmology and Astrophysics (Liaoning) \& College of Sciences, Northeastern University, Shenyang 110819, China}

\author[0000-0002-6754-1448]{Denis Tramonte}
\affiliation{Department of Physics, Xi'an Jiaotong-Liverpool University, 111 Ren'ai Road, Suzhou Dushu Lake Science and Education Innovation District, Suzhou Industrial Park, Suzhou 215123, People's Republic of China}

\author[0000-0001-8075-0909]{Furen Deng}
\affiliation{National Astronomical Observatories, Chinese Academy of Sciences, Beijing 100101, China}
\affiliation{School of Astronomy and Space Science, University of Chinese Academy of Sciences, Beijing 100049, China}
\affiliation{Key Laboratory of Radio Astronomy and Technology, Chinese Academy of Sciences, A20 Datun Road, Chaoyang District, Beijing 100101, China}
\affiliation{Institute of Astronomy, University of Cambridge, Madingley Road, Cambridge, CB3 0HA, UK\\}

\author[0009-0008-7631-7991]{Jiaxin Wang}
\affiliation{Key Laboratory of Cosmology and Astrophysics (Liaoning) \& College of Sciences, Northeastern University, Shenyang 110819, China}

\author[0000-0003-0631-568X]{Yougang Wang}
\affiliation{National Astronomical Observatories, Chinese Academy of Sciences, Beijing 100101, China}
\affiliation{School of Astronomy and Space Science, University of Chinese Academy of Sciences, Beijing 100049, China}
\affiliation{Key Laboratory of Radio Astronomy and Technology, Chinese Academy of Sciences, A20 Datun Road, Chaoyang District, Beijing 100101, China}

\author[0000-0002-6029-1933]{Xin Zhang}
\affiliation{Key Laboratory of Cosmology and Astrophysics (Liaoning) \& College of Sciences, Northeastern University, Shenyang 110819, China}
\affiliation{National Frontiers Science Center for Industrial Intelligence and Systems Optimization, Northeastern University, Shenyang 110819, China}
\affiliation{Key Laboratory of Data Analytics and Optimization for Smart Industry (Ministry of Education), Northeastern University, Shenyang 110819, China}

\author[0000-0001-6475-8863]{Xuelei Chen}
\affiliation{National Astronomical Observatories, Chinese Academy of Sciences, Beijing 100101, China}
\affiliation{Key Laboratory of Cosmology and Astrophysics (Liaoning) \& College of Sciences, Northeastern University, Shenyang 110819, China}
\affiliation{Key Laboratory of Radio Astronomy and Technology, Chinese Academy of Sciences, A20 Datun Road, Chaoyang District, Beijing 100101, China}
\affiliation{School of Astronomy and Space Science, University of Chinese Academy of Sciences, Beijing 100049, China}

\date{Accepted XXX. Received YYY; in original form ZZZ}

\begin{abstract}
Filaments stand as pivotal structures within the cosmic web. 
However, direct detection of the cold gas content of the filaments remains challenging due to its inherently low brightness temperature.
With the TNG hydrodynamical simulations, we demonstrate the effectiveness of isolating 
faint filament \hi signal from the FAST \hi intensity mapping (IM) survey through pairwise stacking of galaxies, which yields an average \hi filament signal amplitude of $\sim 0.29\ {\mu{\rm K}}$ at $z\simeq 0.1$. 
However, our simulations reveal a non-negligible contribution from \hi-rich galaxies within or near the filaments. Particularly, the faint galaxies dominantly contribute to the extra filament \hi signal. Our simulation also shows that the measurement uncertainty is produced by both thermal noise and background variation caused by brightness leakage from surrounding random galaxies. 
Given a fixed total observation time, a wide-field \hi IM survey, which includes a large number of galaxy pairs, can simultaneously reduce thermal noise to below the filament signal level and minimize background variation to a negligible level.
Through the end-to-end simulation, this work demonstrates the critical role of the galaxy pairwise stacking method in future filament \hi detection, 
outlining a road map for filament HI detection in the next-generation \hi IM surveys.
\end{abstract}


\section{Introduction}
At large scales (above $\sim10\,\text{Mpc}$), the distribution of galaxies (and dark matter) shows 
an intricate multi-scale interconnected network which is the so-called \textit{cosmic web} 
\citep{1996Natur.380..603B}, consisting of nodes (dense regions typically hosting clusters of galaxies), 
long filaments (connecting nodes), flattened sheets (or walls), and vast low-density voids.
This striking pattern, which has been found to repeat throughout the observable Universe, was formed in 
the late Universe from the non-linear growth of perturbations in the energy density of the cosmic fluid, thus becoming an important benchmark for testing cosmological models. 

The existence of the cosmic web was suggested by early attempts to map the nearby cosmos 
in galaxy redshift surveys \citep{1978ApJ...222..784G,1978MNRAS.185..357J,1986ApJ...302L...1D,1989Sci...246..897G,1996ApJ...470..172S}, 
and has been confirmed many times by large galaxy redshift surveys, such as the 
Sloan Digital Sky Survey (SDSS, \citealt{2000AJ....120.1579Y,2004ApJ...606..702T}), 
the two-degree-Field Galaxy Redshift Survey (2dFGRS, \citealt{2003astro.ph..6581C}), the VIMOS VLT Deep Survey (VVDS, \citealt{2005A&A...439..845L}), 
the Cosmic Evolution Surveys \citep{2007ApJS..172....1S,2016ApJS..224...24L}, 
the six-degree-Field Galaxy Survey (6dFGS, \citealt{2009MNRAS.399..683J}), 
the Galaxy and Mass Assembly survey (GAMA, \citealt{2011MNRAS.413..971D}), 
the Two Micron All-Sky Survey (2MASS, \citealt{2012ApJS..199...26H}), 
the VIMOS Public Extragalactic Redshift Survey (VIPERS, \citealt{2014A&A...566A.108G}), 
and the SAMI Galaxy Survey \citep{2015MNRAS.447.2857B}.
This large-scale structure (LSS) in the distribution of matter has also been predicted by 
cosmological N-body simulations 
(e.g. \citealt{2005Natur.435..629S,2014MNRAS.444.1518V,2015MNRAS.446..521S}). 

The most prominent and defining features of the cosmic web are the filaments. 
The milestone analysis of \cite{2014MNRAS.441.2923C} and, more recently, \cite{2019MNRAS.487.1607G} 
have shown that most of the volume of the Universe is occupied by voids ($\sim 76\%$), 
followed by walls and filaments, with the nodes only occupying a tiny volume fraction 
($\sim 0.002\%$). 
As for the mass, most of it is contained in filaments: these structures host $\sim 50\%$ of the dark matter
and gas mass, and $\sim 82\%$ of the stellar mass in the Universe.
These fractions show that the study of matter distribution at the largest scales is inevitably tied to that of cosmic filaments.
 
The relevance of filaments in cosmic web studies is varied. First of all, filaments appear 
to act as transport channels along which the diffuse gas 
and galaxies get funneled into the higher density cluster regions \citep{1993ApJ...418..544V,2004ApJ...603....7K},
and which define the connecting structures between higher density complexes \citep{1996Natur.380..603B,2005MNRAS.359..272C,2010MNRAS.408.2163A}; 
filaments are also thought to torque dark matter halos to align their spin axes \citep{2007MNRAS.381...41H,2007MNRAS.375..489H,2009MNRAS.398.1742H}. Besides, filaments produce their own deep potential wells, thus 
giving rise to a gravitational lensing signal at the largest scales, which a few authors claimed to have detected through weak lensing analyses
(e.g. \citealt{2005A&A...440..453D,2007Natur.445..286M}). 
Numerical simulations, however, predict that structures along the line of sight should produce a shear signal 
comparable to that of the target filaments \citep{2006MNRAS.370..656D}, so that the evidence remains 
far from conclusive. 
Finally, the formation of filaments is accompanied by gravitational heating, 
which gradually raises the temperature of the intergalactic medium (IGM) over time and produces 
the so-called warm-hot intergalactic medium (WHIM) by $z = 0$ (e.g. \citealt{1999ApJ...514....1C}). 
In a few cases, the gaseous WHIM residing in filaments was detected in X-ray emission 
\citep{2008A&A...482L..29W,2011MNRAS.415.1961F} and absorption \citep{2009ApJ...695.1351B,2010ApJ...714.1715F}.
On a similar note, the Sunyaev–Zel’dovich (SZ) effect produced by the WHIM has been used to search for 
the imprint of missing baryons
\citep{2007ARA&A..45..221B,2014ApJ...789...55U,2014PhRvD..89b3508V,
2015ApJ...806..113G,2015PhRvL.115s1301H,2015JCAP...09..046M,
2019MNRAS.483..223T,2019A&A...624A..48D},
as this diffuse gaseous component has traditionally been quite challenging to observe due to its low 
density and relatively low temperature. 

In the past two decades, several studies were devoted to tracing the cosmic web using 
the distribution of diffuse gas, particularly of hydrogen due to its high cosmic abundance.
At higher redshifts, evidence for filamentary structures was found in the spectra of 
background sources due to the absorption of \la photons by the neutral IGM along the 
line of sight (e.g. \citealt{2014A&A...572A..31F}).
For low-redshift, however, the \la line falls in the UV range and is therefore blocked by the atmosphere.
Alternatively, the 21-cm emission from the spin-flip hyperfine transition in the 
neutral hydrogen (\hi) ground state \citep{2012RPPh...75h6901P}, lies in the microwave range 
which enables its detection using ground-based facilities. \cite{2009MNRAS.394L...6P} first showed 
the feasibility of tracing cosmic structures using \hi intensity maps.
\cite{2010arXiv1007.3709C} detected the LSS at a redshift of $z\sim 0.8$ 
by cross-correlating 21-cm intensity maps obtained from the Green Bank Telescope (GBT) 
with data from the Deep Extragalactic Evolutionary Probe 2 (DEEP2, \citealt{2003SPIE.4834..161D}) survey.
The same group followed this work up with a more significant detection using  
WiggleZ galaxy survey data \citep{2013ApJ...763L..20M}. 
\cite{2018MNRAS.476.3382A} used the \hi intensity maps acquired from the Parkes radio telescope 
and cross-correlated them with galaxy maps from the 2dFGRS.
 \hi intensity mapping (IM) survey data made significant progress with the newly 
built MeerKAT, e.g. \cite{2023MNRAS.518.6262C} reported a $7.7\sigma$ detection 
of the cross-correlation power spectrum between MeerKAT \hi map and the 
overlapping WiggleZ galaxy catalog, and \cite{2023arXiv230111943P} reported 
the first detection of \hi auto-power spectrum on Mpc scales. \add{Besides, the Canadian Hydrogen Intensity Mapping Experiment (CHIME) performed the first detection of cosmological 21-cm signal made by an interferometer \citep{2023ApJ...947...16A}.}

Other works explored the possibility of using neutral hydrogen to trace the cosmic web. 
\cite{2014MNRAS.444.2236T} explored the possibility of using the \hi line 
to directly observe IGM filaments, while \cite{2017PASJ...69...73H} performed a similar exercise, 
but for the WHIM.
The feasibility of a direct detection of \hi emission from the filamentary gas has 
been explored in \cite{2017MNRAS.468..857K}.
\cite{2019MNRAS.489..385T} employed the same dataset used by \cite{2018MNRAS.476.3382A}, 
but adopted a methodology very similar to the one used by \cite{2019MNRAS.483..223T} 
and \cite{2019A&A...624A..48D}, 
based on stacking galaxy pairs to enhance the signal coming from the large-scale filaments.
However, due to the limitation in sensitivity of Parkes data, 
no compelling detection of an \hi signal from the filament was obtained.

The newly built Five-hundred-meter Aperture Spherical radio Telescope
\citep[FAST,][]{2011IJMPD..20..989N,2016RaSc...51.1060L,2020RAA....20...64J}
is currently the most sensitive radio telescope in the world.
Benefiting from the 19-feed L-band receiver system, FAST is an ideal tool to 
conduct both \hi galaxy and intensity mapping surveys \citep{2020MNRAS.493.5854H}. 
The FAST All Sky \hi survey (FASHI) is one of the FAST \hi galaxy surveys, 
targeting the full sky area within the FAST field of view \citep{2024SCPMA..6719511Z}.
The Commensal Radio Astronomy FasT Survey \citep[CRAFTS;][]{2018IMMag..19..112L}
is simultaneously conducting the surveys for transit searching, 
the Galactic HI survey, and the extragalactic HI galaxies survey
\citep{2019SCPMA..6259506Z}. 
The FAst neuTral HydrOgen intensity Mapping ExpeRiment (FATHOMER),
which is an \hiim survey experiment targeting a cosmological distance, 
was also carried out and the primary data analysis pipeline has been developed \citep{2023ApJ...954..139L}.

In this work, we perform an end-to-end simulation analysis on detecting filamentary \hi 
using the galaxy pairwise stacking for the FAST \hiim drift scan surveys.
The \hi sky map, as well as the corresponding galaxy catalog, are simulated using the 
IllustrisTNG simulation datasets \citep[TNG,][]{2019ComAC...6....2N}.
The paper is organized as follows:
In \refsc{sec:simulation} we provide a brief description of the TNG dataset and the 
\hi cube simulation methods used in this work.
In \refsc{sec:method} we introduce the galaxy pairwise stacking method.
The results and discussion are presented in \refsc{sec:results}.
In \refsc{sec:conclusion} we summarize and conclude. 
We use the standard ${\rm \Lambda}$CDM cosmology model with the cosmology parameters
constrained with {\it Planck} power spectra, in combination with lensing reconstruction and
external data \citep[Table 4 in ][]{2016A&A...594A..13P}, i.e. 
$h=0.6774$, $\Omega_{\rm m}=0.3089$, $\Omega_{\rm \Lambda}=0.6911$, 
which are consistent with the parameters used in the TNG simulation.

\section{Simulation Data}
\label{sec:simulation}

\subsection{TNG simulation snapshot}

\begin{table}
    \begin{center}
    \caption{The physical and numerical parameters of TNG50, TNG100 and TNG300.}\label{tab:tng}
    {\scriptsize
    \begin{tabular}{lccc} 
        \hline
        Run$^\dag$                          & TNG50-1  & TNG100-1 & TNG300-1 \\ \hline
        Volume $[\rm cMpc^3]$               & $51.7^3$ & $106.5^3$ & $302.6^3$ \\
        $L_{\rm box}, [{\rm cMpc}/h]$       & $35$ & $75$ & $205$ \\
        $N_{\rm GAS,DM}$                    & $2160^3$ & $1820^3$ & $2500^3$ \\
        $N_{\rm Tracer}$                    & $1\times2160^3$ & $2\times1820^3$ & $1\times2500^3$  \\
        $m_{\rm baryon}, [{\rm M}\odot/h]$  & $5.7\times10^4$ & $9.4\times10^5$ & $7.6\times10^6$ \\
        $m_{\rm DM}, [{\rm M}\odot/h]$      & $3.1\times10^5$ &  $5.1\times10^6$ & $4.0\times10^7$ \\
		\hline
    \end{tabular}}
    \end{center}
    {\noindent
    $\dag$ We adopt the flagship run of the TNG simulation for each physical box size.
    }
\end{table}

The TNG project \citep{2018MNRAS.475..624N,2018MNRAS.475..676S,2018MNRAS.480.5113M,2018MNRAS.477.1206N} 
includes a suite of large-volume, cosmological, 
gravo-magnetohydrodynamical simulations run with the moving-mesh code \textsc{Arepo} \citep{2010ARA&A..48..391S}. 
The simulations include a comprehensive model for galaxy formation physics, which can realistically 
trace the formation and evolution of galaxies over cosmic time
\citep{2018MNRAS.479.4056W,2018MNRAS.473.4077P,2018MNRAS.475..648P}.
The TNG project comprises three simulation volumes: TNG50, TNG100, and TNG300, 
where the names represent the periodic length of each box in $\rm Mpc$ units. 
The physical and numerical parameters of each TNG simulation volume are listed
in \reftb{tab:tng}.
The simulated \hi sky map in this work is based on the \hiim drift scan cosmic survey, 
which requires a sky area over a few hundred square degrees, e.g. $\sim 200 \deg^2$ for the current
pilot survey \citep{2023ApJ...954..139L}.
We also require a suitable mass resolution of the input simulation box to 
provide detailed structural properties of galaxies and small-scale gas phenomena.
TNG100 provides an ideal balance of volume and resolution, particularly for 
intermediate mass halos \citep{2019ComAC...6....2N}.
We take a snapshot of the TNG100 simulation box at redshift $0.1$, i.e. snapshot-091, which 
corresponds to one of the radio frequency interference (RFI) free bands
\citep{2023ApJ...954..139L}.

\subsection{\hi sky map construction}

\begin{figure}
    \centering
    \includegraphics[width=0.5\textwidth]{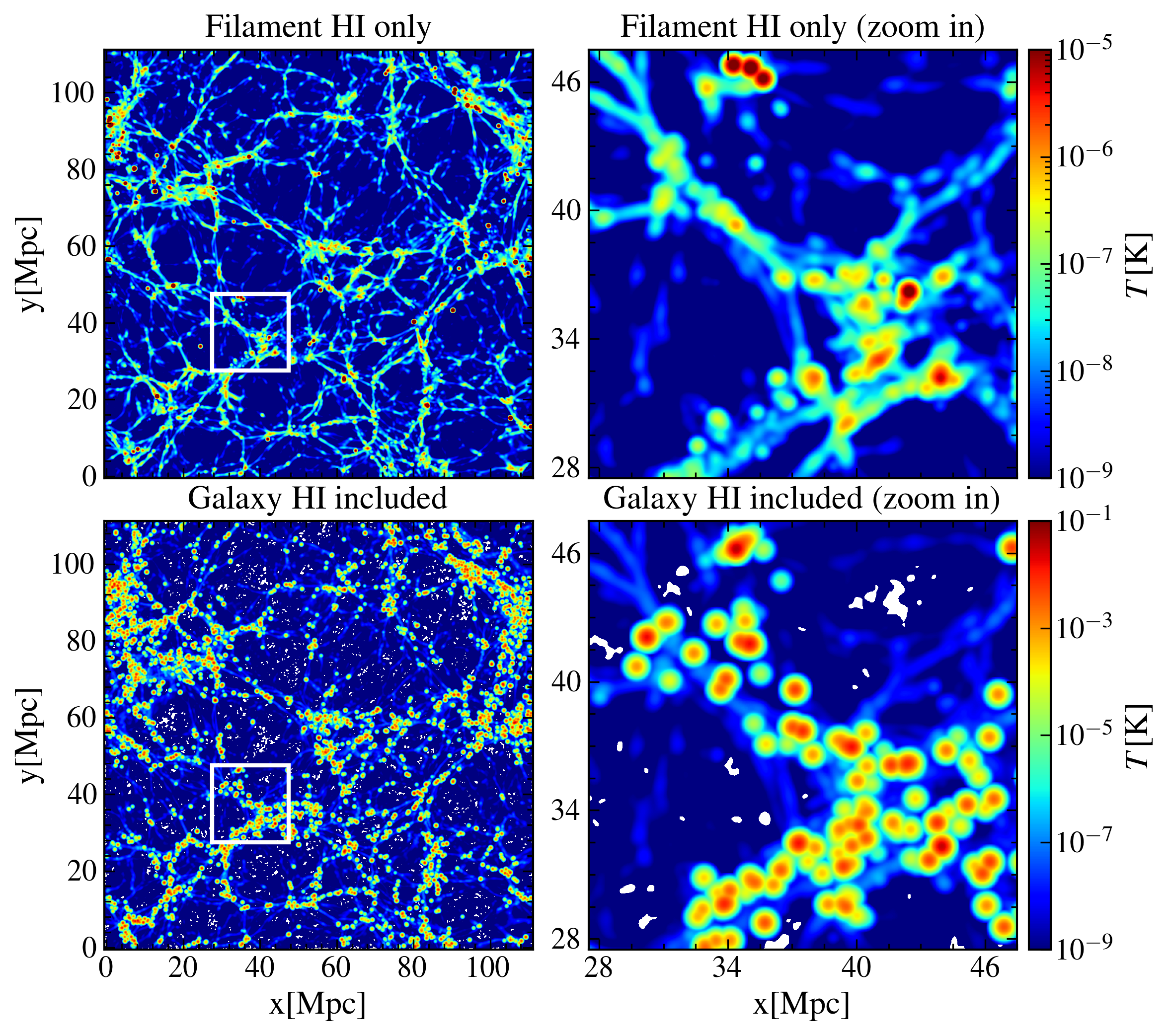}
    \caption{A slice of the construct \hi sky maps at $1.29\, {\rm GHz}$. The left column shows the full maps with 
    the size of $110\,{\rm Mpc}$ and the right column shows a zoom-in area (a size of $20\, {\rm Mpc}$) as indicated with the white square. 
    The top panels show the maps with only the filament \hi signal while the bottom ones are with galaxy \hi included. 
    All the maps are smoothed with the FAST beam as detailed in \refsc{sec: ob effects}.}
    \label{fig: illu_map}
\end{figure}

The \hi sky map is constructed by projecting the \hi brightness temperature of each TNG simulation particle
onto grids in Right Ascension (R.A.), Declination (Dec.), and frequency axes.
The TNG simulations provide the position of each mass particle in Cartesian coordinates.
We set the $z$-axis as the the line of sight (LoS) direction, and the $x$- and $y$-axes as the R.A. and Dec.
direction, respectively.
The center of the simulation box is set at the LoS comoving distance of redshift $0.1$. 
The comoving LoS distance is then converted into redshift by interpolating the redshift-distance relation,
and the frequency in the observation frame is obtained by:
\begin{equation}\label{eq_freq}
    \nu_{\rm obs} = \frac{\nuhi}{ (1+z)(1 + \vpec/c)} \approx \frac{\nuhi}{ 1 + z + \vpec/c} ,
\end{equation}
where $c$ refers to the speed of light, $\nuhi=1420.406\,{\rm MHz}$ is the \hi rest-frame frequency,
$z$ is the redshift caused by the cosmic expansion, and $\vpec$ is the peculiar velocity of each mass particle.
We neglect the second-order term $z \vpec/c$ as the Doppler redshift caused by peculiar velocity is small.
To simplify the calculation in the rest of the analysis, we do not convert the transversal coordinates X and Y into R.A. and Dec. angles. 
The \hi sky map is finally constructed
with a frequency resolution of $0.1\ \mhz$ and a spatial resolution of $20\ \kpc$. 
The full sky area is $0.01\,{\rm Gpc}^2$, corresponding to about $200\ {\rm\deg^2}$ 
with the simulation box at redshift $0.1$.

The TNG simulations provide the total mass of neutral gas, including both the atomic and molecular 
hydrogen, i.e. \hi and H$_2$. 
In high-density regions where stars form, the gas is modeled as a two-phase medium with an effective equation of state for the calculation of cooling, heating and ionization. 
In low-density regions, the gas is assumed to be in ionization equilibrium with a time-evolving ultraviolet(UV)/X-ray background \citep{2013MNRAS.436.3031V,2017MNRAS.465.3291W} with the self-shielding effect included which allow more gas to remain neutral by reducing the UV background's effects\citep{2013MNRAS.430.2427R}. 
The heating and ionization effect of stellar and AGN feedback is also coupled into the calculation of gas state \citep{2009ApJ...703.1416F,2013MNRAS.436.3031V}. 
For a more detailed discussion, refer to the work of \citet{2018ApJS..238...33D}.
The $\hi/{\rm H}_2$ fraction within the neutral gas is then determined with 
the ${\rm H}_2$ abundance model from \citet{2011ApJ...728...88G}, 
which requires the dust-to-gas ratio and UV field strength.
The dust-to-gas ratio is universally assumed to be proportional to the gas metallicity, 
provided by the TNG simulation for each mass particle.
As for the UV photons, they can be generated from two different processes: 
for mass particles with a positive star-forming rate (SFR), the UV field strength is assumed to be 
proportional to the SFR surface density,  
while for mass particles with an SFR equal to or below zero, the UV field was set to be the 
cosmic UV background \citep{2015MNRAS.452.3815L}.
For more details discussion on the calculation of \hi to ${\rm H}_2$ fraction, refer to the work of
\citet{2022MNRAS.515.5894D}.

The $M_\hi$ is then converted to the \hi brightness temperature $T_{\rm b}$ via, 
\begin{equation}
T_{\rm b} = (1+z)^3 \frac{3c^2 h_{\rm P} A_\hi}{32\pi k_{\rm B} m_{\rm H} \nu_{\hi} D_{\rm L}^2} \frac{M_\hi}{\Omega_{\rm p} \delta \nu},
\end{equation}
where $h_{\rm P}$ is the Planck's constant, $k_{\rm B}$ is the Boltzmann’s constant, 
$A_\hi = 2.86888 \times 10^{-15}\, {\rm s}^{-1}$ is \hi spontaneous emission rate, $\nu_{\hi}$ is the rest-frame 
\hi emission frequency,
$m_{\rm H}=1.673533 \times 10^{-27}\, {\rm kg}$ is the mass of the hydrogen atom,
$\Omega_{\rm p}$ is the solid angle of the \hi sky map pixel, and
$\delta \nu=0.1\,{\rm MHz}$ the frequency resolution of the \hi sky map.

In order to verify the pairwise-stacking method, we construct a sky map with only
the filament \hi signal, which is created by removing all particles associated with galaxies. 
The particle averaged neutral hydrogen fraction, defined as the fraction of \hi mass to the total gas mass, 
is $1.9\times 10^{-4}$ for the filament-only sky map.
While the voxel averaged \hi column density (popping out zero value voxels) is $6.6\times 10^{12}\,{\rm cm^{-2}}$ for 
the filament-only sky map.
In addition, we also constructed a normal sky map, which has the galaxy \hi included.
\reffg{fig: illu_map} shows a frequency slice of the two sky maps.
Obviously, the filament \hi only map has a much lower brightness temperature
compared to the galaxy-included map, which indicates that a major portion of \hi
is located within the galaxies. However, even with the filament \hi only map, 
there is still a significant amount of \hi located close to the filament node,
which might be due to the \hi resides in the outer range of large galaxies in the cluster.

\subsection{Observation effects}\label{sec: ob effects}

Beam-smearing and systematic noise are the primary observational effects considered in this work.
We assume a Gaussian beam model and include the beam size evolution with the redshift as in
\citep{2020MNRAS.493.5854H}:
\begin{equation}\label{eq_FASTbeam}
    \theta_{\rm B} = 1.22\times \frac{0.21\ {\rm m}}{300\ {\rm m}}(1+z) = 2.94\ (1+z)\ {\rm arcmin},
\end{equation}
where $\theta_{\rm B}$ represents the full width at half maximum (FWHM) of the Gaussian 
beam profile and $300\ {\rm m}$ is the effective illuminated aperture of FAST.
We ignore the effect of beam sidelobes and the beam shape variation between different
positions in the multi-beam array, as both are secondary effects.
The Gaussian kernel for each frequency/redshift slice is then formed in the physical distance
to match the physical distance units of the simulated sky map.

The observational thermal noise of a single-dish telescope with dual polarization is given by \citep{2017PASA...34...52M}:
\begin{equation}\label{eq_thermalnoise}
    \sigma_{\rm T} = (1+z)^3\frac{{\rm c}^2 \tsys}{\nuhi^2 \aeff \ob}\frac{1}{\sqrt{2\Delta\nu \Delta t}},
\end{equation}
where the system temperature depends on the sky model and the receiver temperature, $\tsys = \tsky + \trec$, and $\ob$ is the solid angle of the primary beam, which is
\begin{equation}\label{eq_beamsolidangel}
    \ob = \frac{\pi\ \theta_{\rm B}^2}{4\ln{2}}
\end{equation}
under our symmetry assumption.

In the FAST \hiim pilot survey \citep{2023ApJ...954..139L}, the feed array was rotated by $23.4^{\degree}$ pointing an area close to the zenith at the telescope site, together with the overlap of 19 beams in horizontal scanning, resulting in an integration time of about $\Delta t =48\ {\rm s}\ \mathrm{per\ beam}$ for one scan \citep{2020MNRAS.493.5854H}.

Following \cite{2020MNRAS.493.5854H}, we used the sky model for high-latitude,
\begin{equation}\label{eq_Tsky}
    \tsky = 2.73 + 25.2\times(0.408/\nu_{\ghz})^{2.75}\ {\rm K},
\end{equation}
and adopted an effective area of $\aeff=50000\ {\rm m^2}$, a receiver temperature of $\trec=20\ {\rm K}$, finally resulting in thermal noise at a level of $\sigma_{\rm T}=9\ \mk$ at redshift $z=0.1$.

Because adding the thermal noise and performing the map stacking are both linear operations, we generated and saved a Gaussian random noise map with $\sigma_{\rm T}=9\ \mk$, which has the same shape as the \hi sky map. This allows us to perform the stacking process for signal and noise separately, and subsequently add the resulting stacked maps to obtain a final stack that includes noise.

\subsection{Galaxy redshift survey catalog construction}

\begin{figure}
    \centering
    \includegraphics[width=0.5\textwidth]{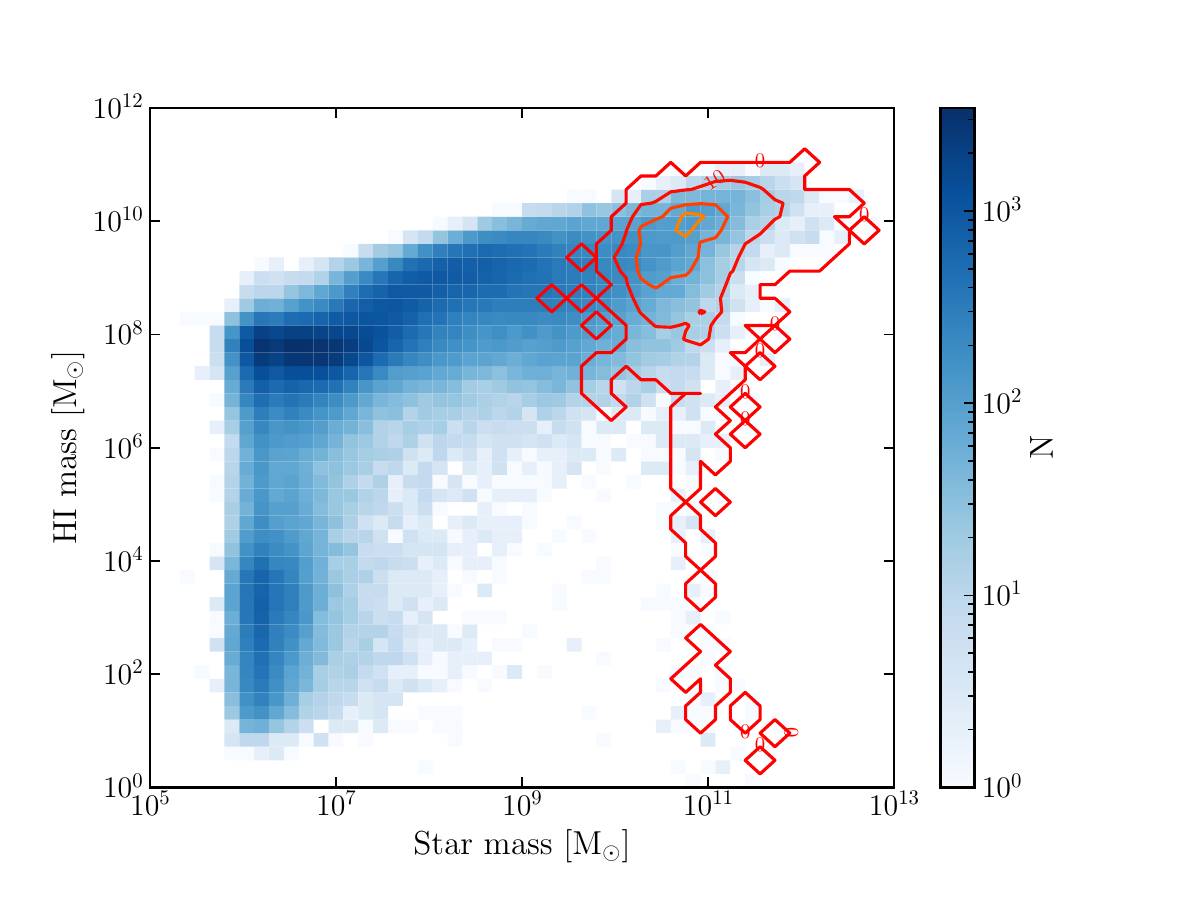}
    \caption{The galaxy samples in the \hi and star mass frame. 
    The histogram presents the distribution of the galaxies from the TNG simulation, while about 4 million galaxies with zero masses are ignored. The red contours denote the selected MGS-like galaxies. }
    \label{fig: Galaxy_mass}
\end{figure}

The stacking analysis requires the positions of galaxies from different observations, 
which can be provided by optical galaxy redshift survey.  
Currently, the FAST \hiim drift scan survey targets the field overlapping with the 
SDSS galaxy surveys and the frequency range is fully covered by the SDSS DR7 Main Galaxy Sample 
\citep[MGS,][]{2002AJ....124.1810S}.

The TNG simulation cube provides the corresponding galaxy catalog via the \textsc{Subfind} algorithm. 
There are more than $4.4$ million subhalos (galaxies) in total within the selected TNG snapshot,
after flagging the fake galaxies mistakenly identified by the algorithm.
To ensure well-defined galaxy samples,
galaxies with both gas and star masses greater than $2\times 10^8 {\rm M}_{\odot}$ \citep{2018ApJS..238...33D} are finally included
as the initial galaxy catalog, which contains $\numprint{37282}$ objects.

The MGS catalog is flux-limited to $r = 17.77$ with a median redshift of $0.1$. 
We apply the $r$-band Petrosian apparent magnitude cutoff of $r \le 17.77$ to the initial galaxy catalog
and form an MGS-like catalog.
The simulated MGS-like catalog includes $\numprint{3526}$ galaxies.
The galaxy samples are illustrated in \reffg{fig: Galaxy_mass} on a \hi mass vs star mass plot.
The distribution of initial galaxies is represented with the blue colormap, 
and the selected sample of MGS-like galaxies is represented by red contours.

The full MGS catalog spans a sky area of $\sim \numprint{7356}\,{\rm deg}^2$ 
\citep{2015MNRAS.449..835R}. Due to the limited TNG box size, we only
simulate a sky field of $\sim 200\,{\deg}^2$ in a narrow redshift bin $0.08<z<0.12$.
The volume number density of the MGS-like catalog is 
$3526/(75\ \hmpc)^3\approx 8.36\times 10^{-3}\ (\hmpc)^{-3}$, which is consistent 
with that derived by querying the MGS galaxy catalog from the SDSS database
within the same redshift range.

\section{Galaxy Pairwise-Stacking Methods}\label{sec:method}

\begin{figure*}
    \centering
    \includegraphics[width=\textwidth]{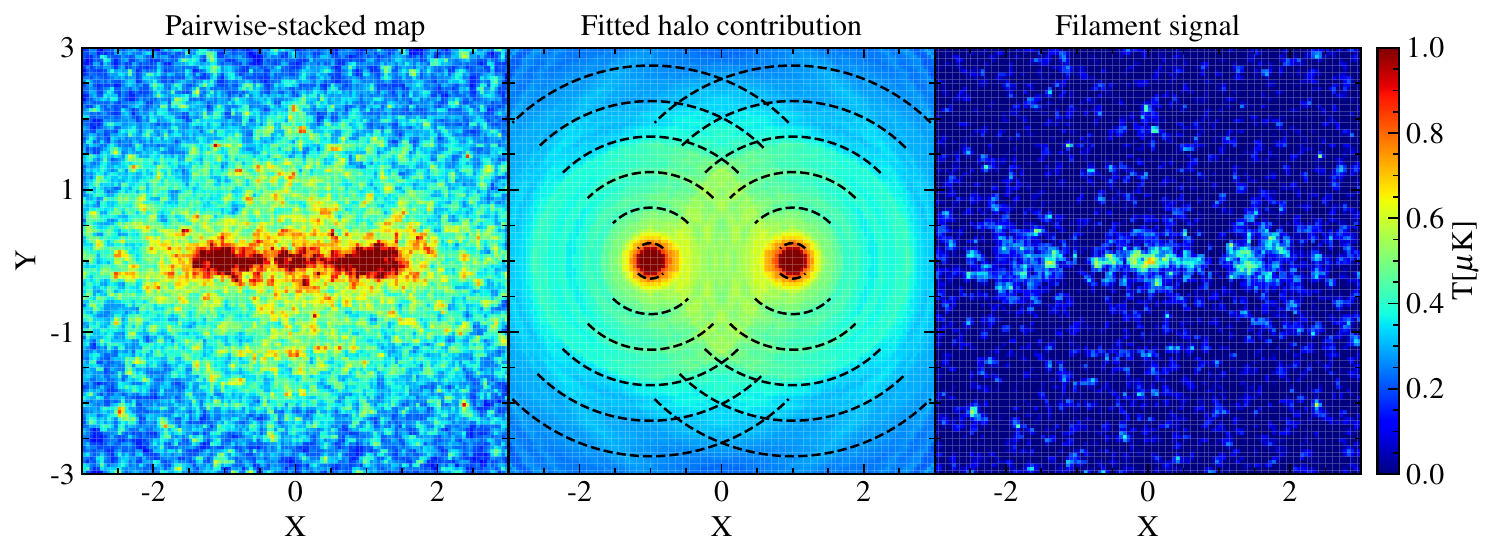}
	\caption{Result of stacking the \mgspairs on the filament \hi only map, showing the final stack map (\textit{left}), the halo contribution best-fit map (\textit{middle}) and the filament residual map (\textit{right}). The black dashed lines in the center panel denote the area used to fit the halo contribution.}
    \label{fig: illu_results}
\end{figure*}

\begin{figure}
    \centering
    \includegraphics[width=0.5\textwidth]{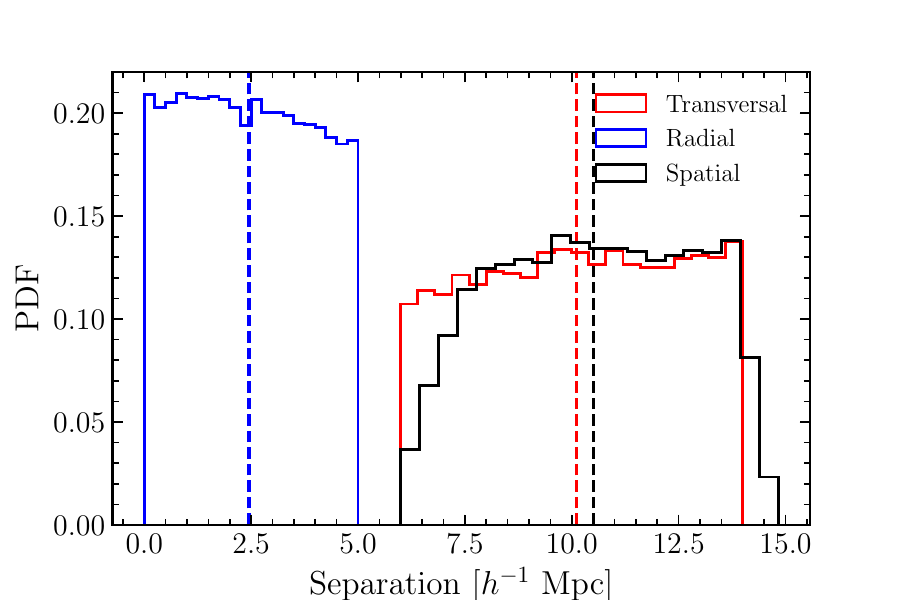}
    \caption{
    The probability density of the separation distances of the 
    selected MGS-like galaxy pairs. 
    The red, blue, and black lines represent the separation distances in the transversal, 
    radial, and spatial directions, respectively. 
    The dashed vertical lines represent the mean separation distances. }
    \label{fig: MGSpairs_sep}
\end{figure}

We adopt the method of stacking galaxy pairs to enhance the filamentary \hi signal-to-noise 
ratio. The galaxy pairwise stacking to search for \hi in filament 
has already been adopted in the literature \citep[e.g.][and the references therein]{2019MNRAS.489..385T}.
Here is a brief summary of the process.

Such a blind stacking method relies on the assumption that 
filaments align with the line connecting a pair of galaxies. 
We use the galaxies as proxies for the cluster positions.
According to the simulation-based study in \cite{2005MNRAS.359..272C}, 
the probability of finding a filament between clusters 
is about $60\%$ when their separation is within $5\sim15\ \hmpc$.
The cluster pairs with separation lower than $5\ \hmpc$ are connected by a filament
with a $\sim 100\%$ probability; nonetheless, such filaments are challenging to distinguish from the merging outskirts of the two clusters.  
Therefore, we chose the galaxy pairs with transversal separation distance 
within 6 and 14 $\hmpc$ and radial separation distances below $5\ \hmpc$
\citep{2019A&A...624A..48D,2019MNRAS.489..385T}.
The lowest limit in the transversal separation prevents contamination from the 
potential projection of two halos along the LoS. It also ensures that we are considering a pair of galaxies belonging to different clusters, 
which helps us to target the large-scale filament structures linking the clusters.

There are $\numprint{76266}$ galaxy pairs selected in this work,
which is about $1.23\%$ of the total galaxy pairs.
\add{The probability density of the separation distance of those selected pairs is shown in the \reffg{fig: MGSpairs_sep}.} 
It slightly increases with their lateral transversal separations, while it slightly decreases with their lateral radial separations.
The mean values are approximately $\sim 10.1\ h^{-1}{\rm Mpc}$ in the transversal direction and $\sim 2.5\ h^{-1}{\rm Mpc}$ in the radial direction.

\citet{2005MNRAS.359..272C} demonstrated that nearly half of the filaments are not strictly aligned with the line connecting the endpoint cluster centers, particularly those extending beyond $15\,h^{-1}{\rm Mpc}$. Our pairwise selection criterion, which limits the separation to less than approximately $15\,h^{-1}{\rm Mpc}$, effectively reduces the number of non-straight filaments in the sample. Nonetheless, the presence of some non-straight filaments could still diminish the signal-to-noise ratio (SNR) in the stacked filament signal.

\begin{figure}
    \centering
    \includegraphics[width=0.45\textwidth]{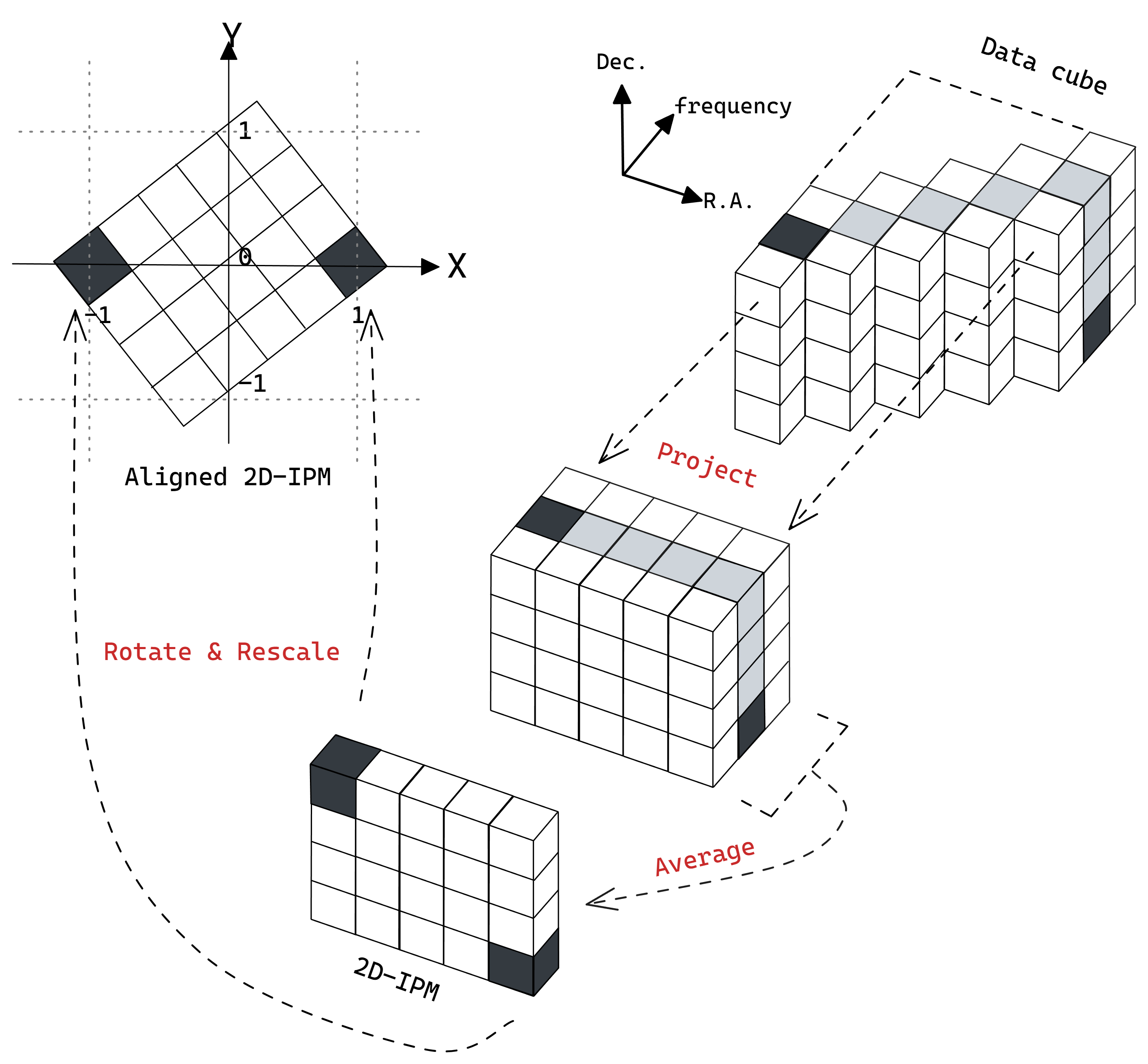}
    \caption{
    \add{Sketch illustrating the construction of the aligned 2D-IPM.  
    The two black voxels represent the positions of galaxy pairs, 
    the voxels in light grey are the relevant data voxels in the diagonal plane,
    and the white voxels represent the data in the front and rear planes.
    The data in the rest of the voxels of the cuboid are ignored to gain better illustration performance.}
    }
    \label{fig: stack_illu}
\end{figure}

\add{The pairwise stacking procedure is illustrated in \reffg{fig: stack_illu} 
and the detailed procedure is listed below:}
\begin{enumerate}
\item {\it Construct the 2D individual pair map (2D-IPM).} 
\add{We define a cuboid using a pair of galaxies as its diagonal vertices, aligning it along the R.A., Dec., 
and LoS axes. The diagonal plane of the cuboid is chosen to be parallel to either the R.A. or Dec. axis, 
depending on which has the shorter angular separation between the galaxy pair. 
The voxels within this diagonal plane are then projected along the LoS to construct one 2D-IPM 
(as demonstrated in \reffg{fig: stack_illu}).
For multiple-slice cases, the front and rear slices are also projected and averaged to obtain the 2D-IPM.}

\item {\it Construct the aligned 2D-IPM.}
After obtaining a 2D-IPM, we rotate and rescale it to a dimensionless $X$-$Y$ frame 
with the two galaxies placed at the reference positions $(-1,0)$ and $(1,0)$.
To facilitate the subsequent analysis, we only intercepted a square area in the aligned 2D-IPM from -3 to 3 in both the X and Y axes.

\item {\it Construct the 2D pairwise-stacked map (2D-PSM).} 
The first two steps are applied to each galaxy pair to form the associated 2D-IPM. 
The 2D-PSM is then constructed by averaging over all the aligned 2D-IPMs.
The rotation and rescaling performed in the second step ensure that the member galaxies from all pairs coherently fall on 
the same positions. 

\end{enumerate}

The \hi filament pairwise stacking analysis has been applied to the 
Parkes \hiim survey maps \citep{2019MNRAS.489..385T}.
Compared to those Parkes \hiim maps, which have a frequency resolution of $1\,\mhz$, we use maps with a higher frequency resolution of $0.1\,\mhz$.
\add{Considering the spectral broadening caused by filament spin \citep{2021NatAs...5..839W}, 
the number of frequency slices used for stacking may influence our results. 
In the subsequent analysis, we start with the 1-slice case, where only the central frequency 
slice of each galaxy pair is used. This corresponds to the diagonal plane, 
represented by grey voxels in \reffg{fig: stack_illu}. 
We then gradually expand the frequency slice range to include the front and rear planes, 
shown as white voxels in \reffg{fig: stack_illu}, to assess the impact of broader frequency widths.}
Unless otherwise specified, any reference to a stacking scenario pertains to the 1-slice case.
See \refsc{sec: multi_slices} for details about the comparison of different slice cases.

The stacking analysis performed with the Parkes \hiim maps also adopted the noise-inverse 
weighted averaging when constructing the 2D-PSM.
The noise variance for each 2D-IPM is extracted from the corresponding 
noise maps, which are delivered together with the \hi maps \citep{2018MNRAS.476.3382A}.
In this simulation, the pairwise stacking analysis is performed using
both the pure \hi simulation maps and the noise-added maps.
We assume the noise to be observational thermal noise, which
is uniform for each 2D-IPM; therefore, the weighting processing does not affect this simulated data and is ignored in this simulation analysis.

\add{Although the simulated \hi sky map is constructed in redshift space, 
the redshift space distortions (RSD) may have minimal effects on our stacking results, 
as the physical distance between galaxy pairs is rescaled to match the dimensionless $X$-$Y$ frame.}

\add{In addition, the two pair-member galaxies may not reside in the same environment; for instance, 
one may be located at the center of a halo while the other is a satellite galaxy.
This asymmetry could lead to varying contributions to the stacked endpoint halo signal.
To ensure a symmetrical signal assignment at $(-1,0)$ and $(1,0)$ in the final 2D-PSM,
we apply a series of procedures to break unintended symmetries and introduce randomness.
First, during the construction of each aligned 2D-IPM, we randomly swap the positions of the galaxy pair
to minimize systematic differences between $(-1,0)$ and $(1,0)$;
Then, before stacking individual 2D-IPMs, we randomly flip the image along the 
$X$ axis, $Y$ axis, or both to further eliminate the asymmetry systematics 
in the final 2D-PSM.}

\add{The full galaxy pair sample is divided into $153$ sub-samples, each used to construct an individual 2D-PSM. 
These $153$ 2D-PSMs are then averaged to form the final 2D-PSM.
Additionally, each 2D-PSM serves as a jackknife sub-sample, 
where one sub-sample is removed at a time to construct a jackknife estimate. 
This allows us to assess the uncertainty in the halo profile estimation (for details, see \refap{sec: halo_sub}).
The full galaxy pair sample is randomly divided into sub-samples 
to ensure that galaxy pairs sharing the same galaxy are uniformly distributed across the sub-samples.}

\section{Results and Discussions}\label{sec:results}

\subsection{Sky maps with only the filament \hi signal}
We begin with the \hi map that only has the filament \hi signal
to demonstrate the capability of extracting the filament signal from a FAST 
\hiim survey using the pairwise stacking method. We also use the MGS-like 
galaxy catalog as the cluster position proxy for the stacking analysis.

\subsubsection{Subtracting the halo contribution}\label{sec:sub_halo}

The 2D-PSM result using the simulated \hi map and the corresponding MGS-like catalog
is shown in the left panel of \reffg{fig: illu_results}, in units of \hi brightness temperature. 
Obviously, the \hi brightness temperature peaks at the positions $(-1,0)$ and $(1,0)$;
these peaks are produced by the residual \hi residing within the halos, which needs to be removed.
Because of the rotation and rescaling applied to the individual 2D-IPMs in the stacking procedure, 
the halo profiles shown in the final 2D-PSM are the merged contributions from individual 
halos of different sizes. A physical modeling of the profile for these endpoint peaks 
is then unfeasible.  

In previous analyses \citep[e.g.][]{2019MNRAS.483..223T,2019MNRAS.489..385T},
the halo contributions were subtracted by fitting a double circular symmetric profile directly on the 2D-PSM.
To prevent subtracting the \hi filament signal, the profile fitting was 
performed after masking the central area between the two peaks, where
the filament is expected to reside.
In this work, we also consider the possibility that the pair-member galaxies may
not be located at the two ends of the filament but reside in knots, with the filament structure 
 extending further along the $X$-axis in the 2D-PSM.
Hence, we restrict the profile fitting to the region composed of two sectors connected at the two peak positions~\citep{2019A&A...624A..48D}, excluding the central region and the external regions bounded by 
$45\,\deg$ tilted lines, see the central panel in \reffg{fig: illu_results}.

Even in this case, we assume the halo contribution profile to be circular-rotationally symmetric 
and we also assume it to be the same at both peaks.
The resulting best-fit profile, projected onto the 2D map, is shown in the middle panel of 
\reffg{fig: illu_results} and the profile removed result is shown in the right panel.
After subtracting the halo contribution, the mean \hi brightness temperature 
of the 2D-PSM is significantly reduced, and clear filament residuals are shown along the $Y=0$ axis.

\subsubsection{Filament quantification}\label{sec:quantification}

\begin{figure}
    \centering
    \includegraphics[width=0.5\textwidth]{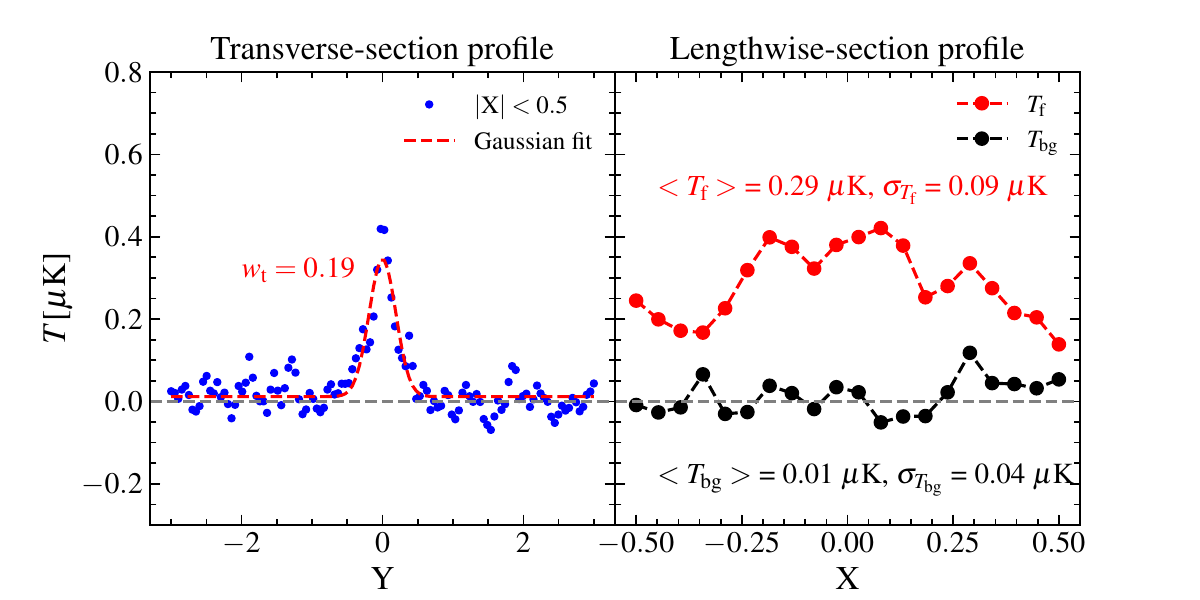}
    \caption{One-dimensional profiles illustrating the filament reconstruction. 
    \textit{Left}: Transverse section filament profiles, with the measured profile shown as a solid blue line, and its Gaussian best fit depicted by a dashed red line. The Gaussian-fitted filament width is displayed in red text.
    \textit{Right}: Lengthwise section profiles, where red markers represent the filament profile and black markers represent the background profile. The mean values and standard deviations of the estimated filament and background level are shown in red and black text, respectively. 
    The gray dashed line indicates $T = 0$.}
    \label{fig: fuzz_profiles}
\end{figure}

In order to quantify the filament brightness temperature, we extract a
1D filament profile along both its transverse and lengthwise sections.
The transverse section profile is obtained by averaging the 2D-PSM 
across the $X$-axis (so along the vertical direction), considering only the pixels satisfying $|X|\le 0.5$ 
to avoid residual contamination from the halos.
The transverse section profile is then fitted with a Gaussian function,
\begin{equation}\label{eq:profile}
T = A \exp\left(-\frac{1}{2}\frac{Y^2}{w_{\rm t}^2}\right) + C,
\end{equation}
where $A$ is the filament peak brightness temperature, 
$w_{\rm t}$ represents the filament transverse section width, and $C$ indicates its offset from zero.
The reconstructed and fitted transverse profiles are illustrated in the left panel of \reffg{fig: fuzz_profiles}.

The lengthwise section profile is defined as the filament brightness temperature averaged along the $Y$-axis, considering only the pixels that satisfy $|Y| \le w_{\rm t}$. 
The lengthwise section profile for $|X| \le 0.5$ is represented by the red markers in the right panels of \reffg{fig: fuzz_profiles}. 
The mean filament signal is estimated as the mean value of this profile, as indicated by the red text in the right panel of \reffg{fig: fuzz_profiles}.

The reference noise level for the filament signal includes both the observational thermal noise (if injected) and the non-symmetrical \hi contribution. To distinguish it from the thermal noise, we refer to this noise level as the background level. More details can be found in \refsc{sec: bglevel}.

The background level is obtained by averaging the 2D-PSM across the $Y$-axis
outside the transverse section profile. 
To maintain the same size of area for estimation, we consider only the pixels
satisfying $3w_{\rm t} < |Y| < 4w_{\rm t}$ and $|X|\le 0.5$.
The resulting background profile is shown with black markers in the right panel of \reffg{fig: fuzz_profiles}. The mean background brightness temperature across the lengthwise section profile, $T_{\rm bg}$, is consistent with zero within the uncertainty level. Therefore, we use the standard deviation of the background region, $\sigma_{T_{\rm bg}}$, to estimate the background level, as indicated by the black text in the right panel of \reffg{fig: fuzz_profiles}.

We take $T_{\rm f} = 0.29\ {\rm \mu K}$ and $\sigma_{T_{\rm bg}} = 0.04\ {\rm \mu K}$ as our estimated signal amplitude and background level, respectively. 
This yields an SNR of $7.25$ without thermal noise injection, where the SNR is calculated using the relation:
\begin{equation}\label{eq:snr} 
\mathrm{SNR} = \frac{T_{\rm f}}{\sigma_{T_{\rm bg}}}. \end{equation}
This result demonstrates the capability of isolating the \hi filaments using the pairwise stacking method.

\subsection{Sky map with galaxy \hi included}

In a more realistic scenario, the \hiim map incorporates contributions from galaxies, which are several orders of magnitude higher than the filament brightness temperature, as illustrated in \reffg{fig: illu_map}.

\subsubsection{The galaxy \hi contribution}\label{sec: galaxy_contributed}

\begin{figure}
    \centering
    \includegraphics[width=0.5\textwidth]{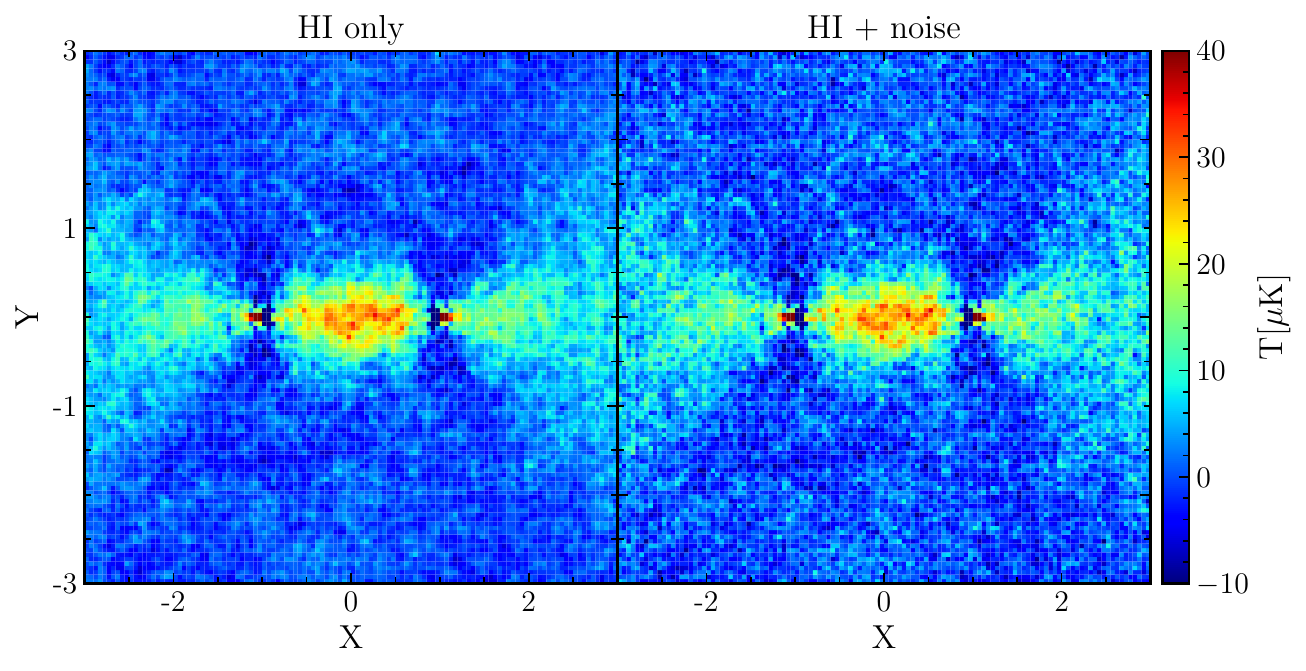}
    \caption{The filament structure reconstructed from the stack of \mgspairs on the galaxy-included \hi map, without masking the galaxies' contributions.
    The left and right panels show the results obtained from using pure \hi signal maps and with added thermal noise, respectively.
	The color scale is adjusted to best show the filament's candy-shaped structure.}
    \label{fig: unmask}
    \includegraphics[width=0.5\textwidth]{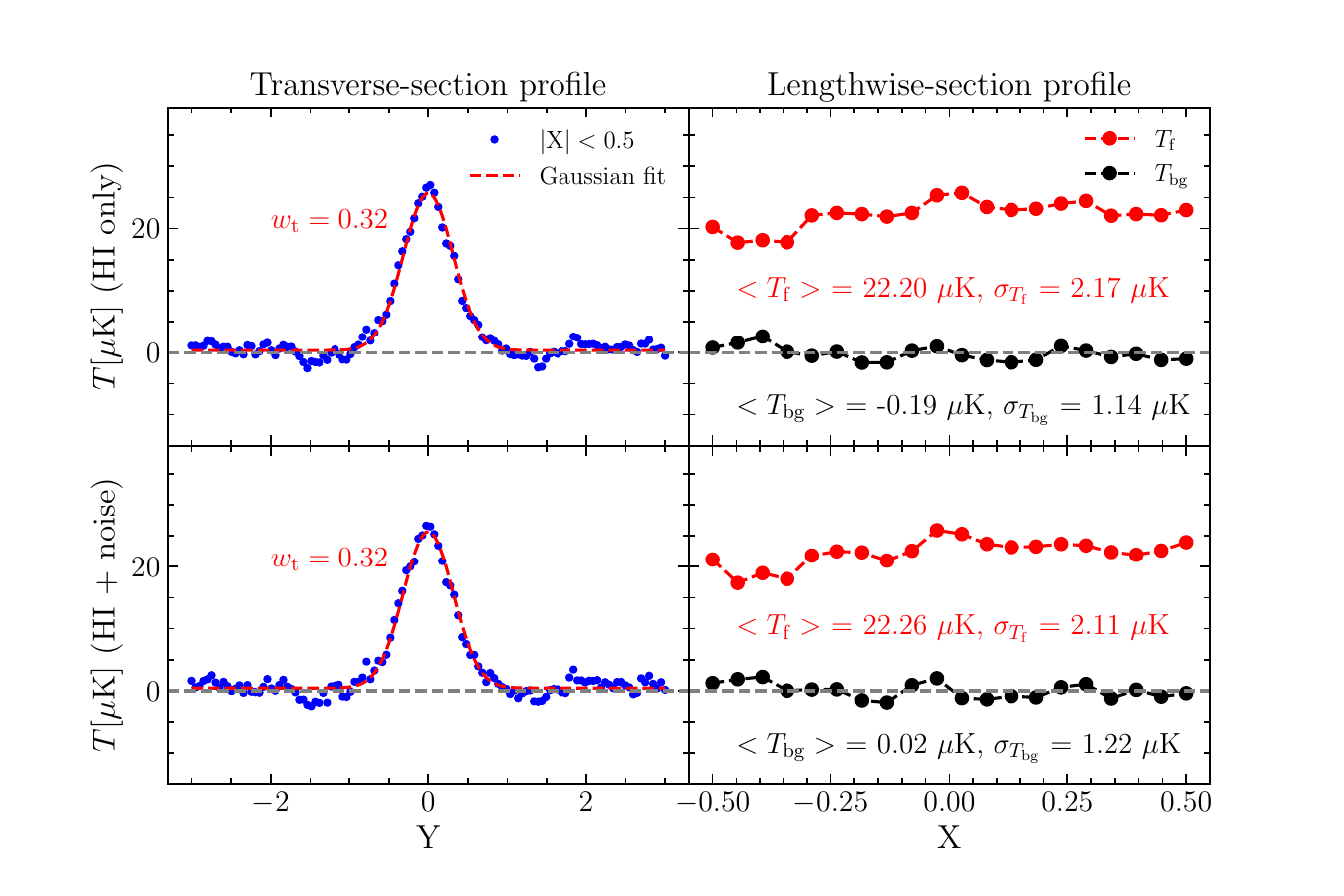}
    \caption{One-dimensional profiles similar to \reffg{fig: fuzz_profiles}, but obtained from the 
    galaxy-included \hi map without masking the galaxies' contributions.
    The top panels show the results obtained using pure \hi signal maps, whereas the
    bottom panels show the results obtained with thermal noise added to the maps.}
    \label{fig: unmask_profiles}
\end{figure}

For comparative analysis, we replicated the aforementioned processes on the galaxy-included \hi map. 
The filament signal of this "Unmasked" result is presented in the left panel of \reffg{fig: unmask}, alongside the thermal noise injection case displayed on the right. 
The results demonstrate a notable increase in filament amplitude and a broadening of the filament distribution, exhibiting a structure reminiscent of a "candy" formation, i.e. a cylinder with fan shapes at the two ends of it. 
Such increases in amplitude and broadening in shape are attributable to the varying contributions of different galaxies, which are highly asymmetrical and luminous.
Furthermore, the thermal noise determines no significant impact on the filament reconstruction.

Through the quantification method introduced above, we obtained the filament profiles in the "Unmasked" case, as shown in \reffg{fig: unmask_profiles}.
The filament width increases from $0.19$ times the scaled length unit to $0.32$ times.
The mean signal and background level increase from $0.29\ {\rm \mu K}$ and $0.04\ {\rm \mu K}$ to $22.20\ {\rm \mu K}$ and $1.14\ {\rm \mu K}$, respectively.
Although the signal amplitude is approximately two orders of magnitude larger, the SNR is only approximately \add{three} times larger. 
This is due to the background level also increasing, \add{to about 28 times larger}, in the case of galaxy contribution.

\subsubsection{Masking galaxy contributions}

\begin{figure}
    \centering
    \includegraphics[width=0.5\textwidth]{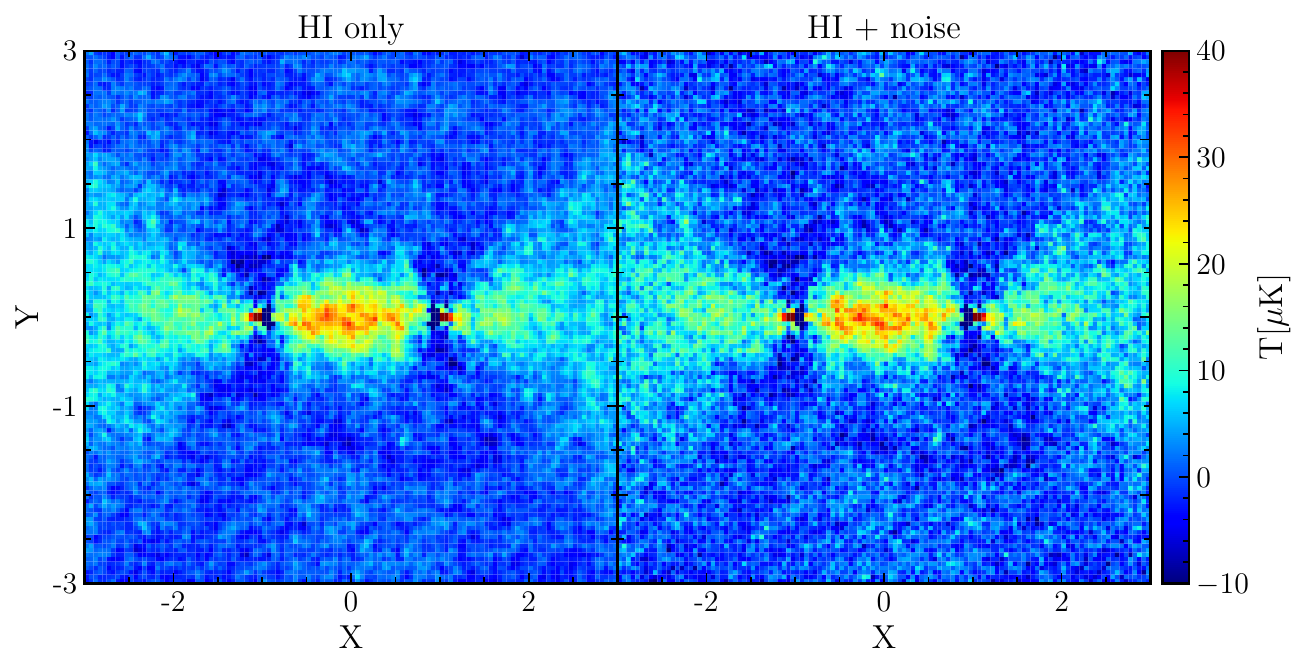}
    \caption{The reconstructed filament structure as in \reffg{fig: unmask}, 
    but obtained after masking the galaxy contribution based on the MGS-like catalog.}
    \label{fig: maskMGS_residual}
    \includegraphics[width=0.5\textwidth]{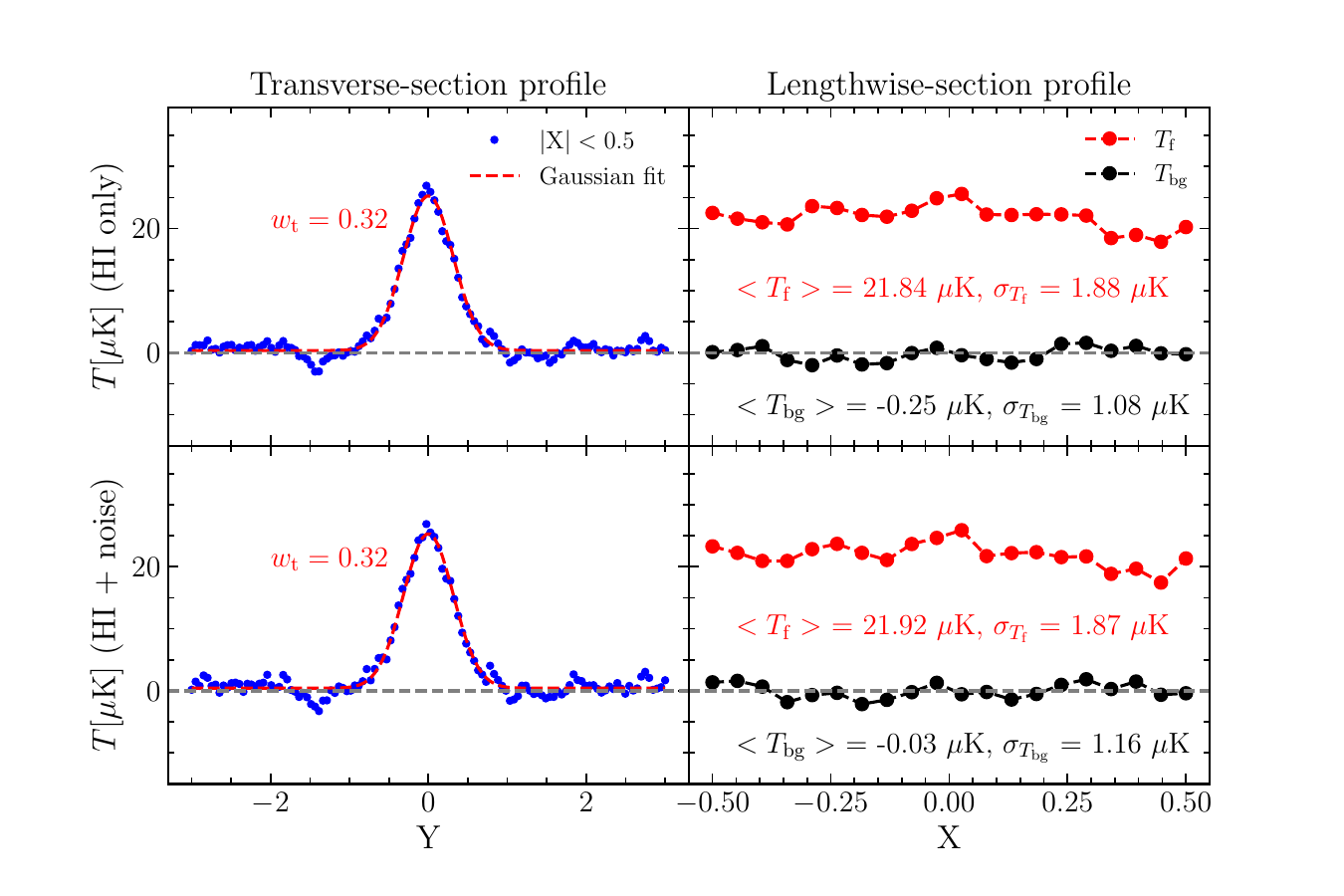}
    \caption{One-dimensional profiles similar to \reffg{fig: unmask_profiles}, 
    but obtained from the galaxy-included \hi map after masking the galaxy contribution based on the MGS-like catalog.
    }
    \label{fig: maskMGS_profiles}
\end{figure}

\begin{figure}
    \centering
    \includegraphics[width=0.5\textwidth]{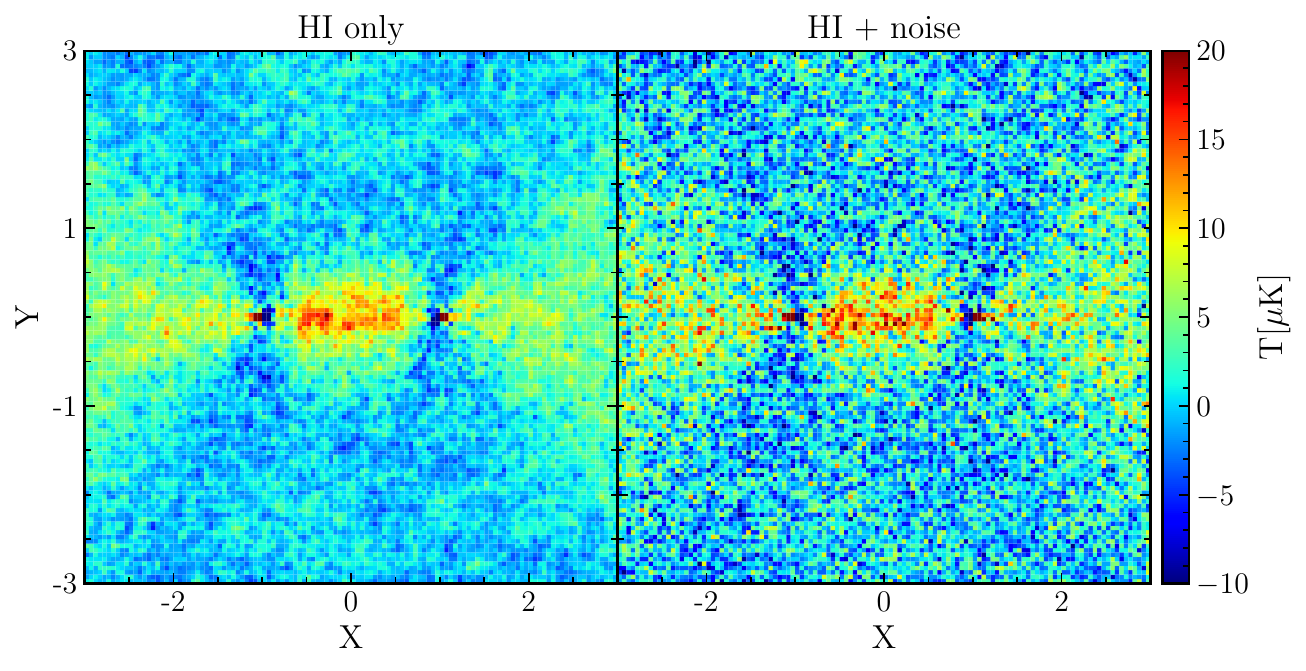}
    \caption{The reconstructed filament structure as in \reffg{fig: unmask}, 
    but obtained after masking the galaxy contribution based on the full galaxy sample.}
    \label{fig: maskAll_residual}
    \includegraphics[width=0.5\textwidth]{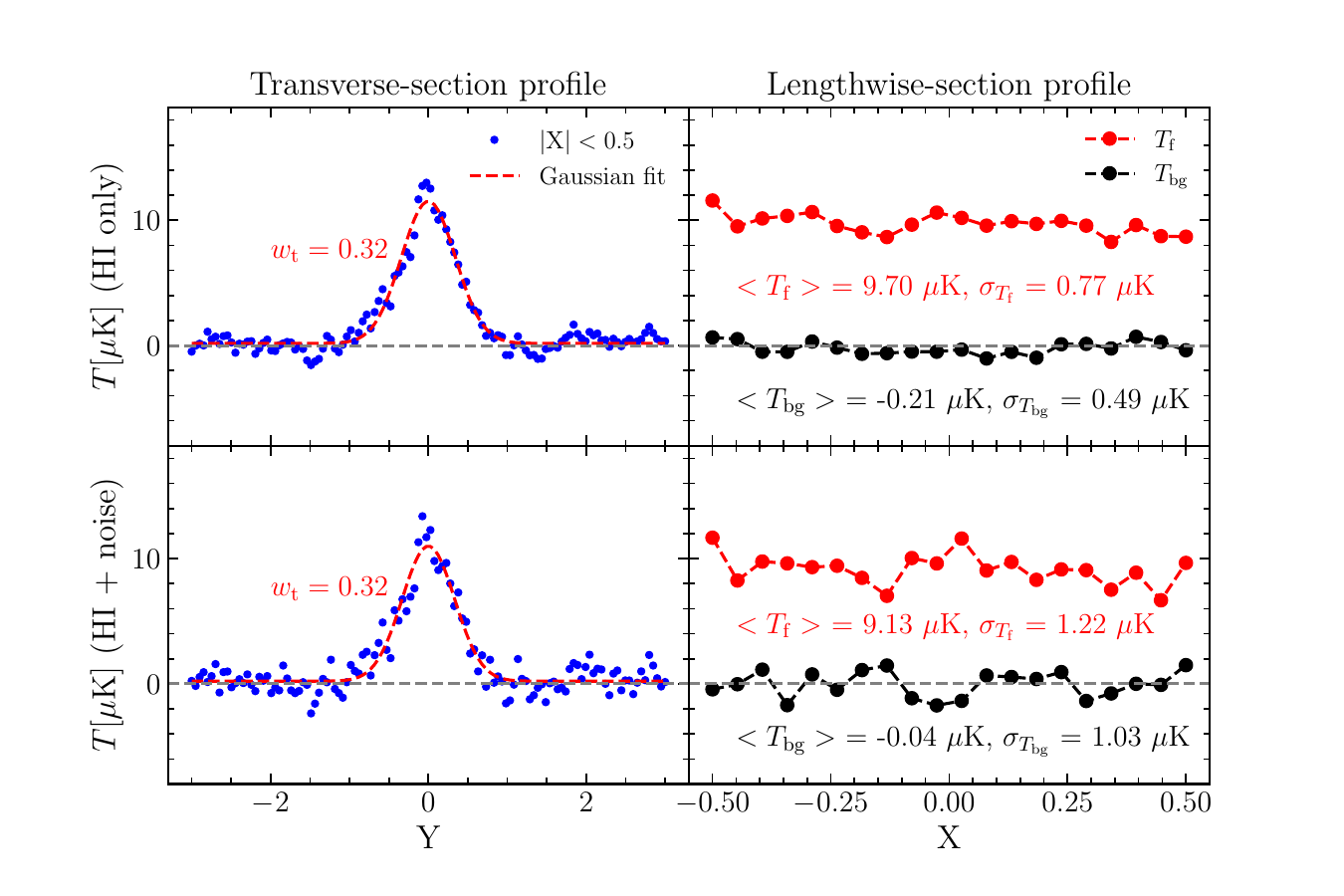}
    \caption{One-dimensional profiles similar to \reffg{fig: unmask_profiles}, 
    but obtained from the galaxy-included \hi map after masking the galaxy contribution based on the full galaxy sample.}
    \label{fig: maskAll_profiles}
\end{figure}

\begin{table}
    \caption{
    Measured filament properties corresponding to different stacking map cases, all obtained within the 
    central frequency slice only. 
    $r$ represents the filament radius defined in \refeq{eq:filamentwidth}, SNR represents the signal-to-noise 
    ratio defined in \refeq{eq:snr}.
    $T_{\rm f}$ and $T_{\rm bg}$ are respectively the filament and background mean brightness temperature. 
    The ``Unmasked'' rows are the results with only the halo contribution removed; 
    the ``Mask MGS'' and ``Mask all'' rows represent the results with masking 
    the contribution from the galaxies in the MGS-like catalog and the full TNG galaxy catalog, respectively.}
    \label{tab: diff mask result compare}
    \centering
    {\small
    \begin{tabular}{lccccc}
        \hline
        & $r$        & $T_{\rm f}$    & $T_{\rm bg}$   & SNR \\
        &[$h^{-1}{\rm Mpc}$] &[${\rm \mu K}$]  &[${\rm \mu K}$]  & \\   
        \hline
        & \multicolumn{4}{c}{filament \hi only} \\ \cline{2-5}
             &0.96 &$0.29\pm0.09$ &$0.01\pm0.04$ &7.25\\
        \hline
        & \multicolumn{4}{c}{galaxy \hi contributed} \\ \cline{2-5}
        Unmasked     &1.61 &$22.20\pm2.17$ &$-0.19\pm1.14$ &19.47\\
        Mask MGS     &1.61 &$21.84\pm1.88$ &$-0.25\pm1.08$ &20.22\\
        Mask all     &1.61 &$ 9.70\pm0.77$ &$-0.21\pm0.49$ &19.80\\
        \hline
        & \multicolumn{4}{c}{galaxy \hi contributed, with noise injection} \\ \cline{2-5}
        Unmasked     &1.61 &$22.26\pm2.11$ &$\,0.02\pm1.22$ &18.25\\
        Mask MGS     &1.61 &$21.92\pm1.87$ &$-0.03\pm1.16$  &18.90\\
        Mask all     &1.61 &$ 9.13\pm1.22$ &$-0.04\pm1.03$  & 8.86\\
        \hline
    \end{tabular}
    }
\end{table}

To exclude the galaxy \hi contributions, \cite{2019MNRAS.490.1415K} efficiently masked the 
signal within a fixed radius $r = 100\ h^{-1}{\rm kpc}$ from each galaxy position. 
The mask radius is larger than the mean galaxy size to avoid signal leakage \citep{1993ApJ...419..515R}.
We adopt the same strategy in our analysis. However, the FWHM beam size of FAST is 
$\sim408\ \kpc$ at a redshift of $0.1$, which
corresponds to a radius size of $\sim 120\,{h}^{-1}{\rm kpc}$.
To simplify the calculation, we assume a constant symmetrical Gaussian beam model 
without beam size variations across frequencies
and adopt a constant mask radius of $\sim 120\,{h}^{-1}{\rm kpc}$.

However, determining the optimal number of masked frequency slices to minimize the contribution of galaxies 
presents a significant challenge, as this contribution varies among different galaxies. 
In the subsequent analysis, we initially focus on the central 1-slice case,
deferring the detailed comparisons to \refsc{sec: multi_slices}.

The coordinates of the masked galaxies are provided from 
the same optical survey catalog as the one used to select the pairwise galaxies.
In this simulation, we adopt the MGS-like catalog for galaxy masking.
The MGS-like galaxy-masked filament pairwise stacking results are shown 
in \reffg{fig: maskMGS_residual}, with the pure \hi result in the 
left panel and noise inserted result in the right panel; 
the corresponding filament profiles are shown in \reffg{fig: maskMGS_profiles}.

In comparison to the previous results obtained without galaxy masking, the reconstructed signal for the "Mask MGS" case exhibits no significant \add{changes in amplitude and also SNR, with $T_{\rm f} = 21.84\,{\rm \mu K}$ and ${\rm SNR}=20.22$.}
The injection of thermal noise also does not substantially impact the reconstruction of the filament, resulting in a comparable signal level detection, although it does lead to a slight decline in SNR.

However, the MGS-like catalog is recognized as an incomplete sample due to its magnitude limit. 
Building on the simulation, we further investigate the scenario where the full galaxy samples are masked. 
Constructing the "full galaxy sample" presents challenges for the TNG simulation, as the accuracy of galaxy identification diminishes near the mass limit. 
The simulation's mass resolution is limited at approximately $10^6\ M_{\odot}$ for baryon particles.
As the mass of a galaxy decreases and approaches this resolution limit, the number of particles within the galaxy also decreases, sometimes reducing to just a few particles. At this stage, the particles fail to adequately represent the galaxy, leading to a higher likelihood of misidentification by galaxy identifier algorithms, e.g. \textsc{subfind} \citep{2001MNRAS.328..726S}.
To address this, we implement an aggressive galaxy-masking solution that includes all galaxies identified in the TNG simulation, irrespective of the mass limit. 
\add{In this test scenario, we deliberately included the extremely low-mass galaxies 
as a source of contamination, which may not fully reflect real astrophysical conditions. 
This was designed as an aggressive test to evaluate the robustness and limitations of 
the stacking method itself.}
The results of the associated filament residuals are presented in \reffg{fig: maskAll_residual}, with the corresponding profiles illustrated in \reffg{fig: maskAll_profiles}.

As depicted in \reffg{fig: maskAll_residual}, the filament's candy-shaped structure becomes more diffuse, and its brightness temperature is significantly reduced; with the introduction of thermal noise, the filament becomes even noisier.
As illustrated in \reffg{fig: maskAll_profiles}, although the filament's transverse section width experiences slight variations, the Gaussian-fitted longitudinal profiles in both scenarios, absence and presence of noise, yields comparable mean filament signal amplitudes of $T_{\rm f} = 9.7\ {\rm \mu K}$ and $T_{\rm f} = 9.13\ {\rm \mu K}$, respectively. 
However, the background levels for these two cases differ, with $T_{\rm bg} = -0.21\pm\, 0.49\ {\rm \mu K}$ in the absence of noise and $T_{\rm bg} = -0.04\pm\, 1.03\ {\rm \mu K}$ when noise is injected. 
This disparity indicates a significant reduction in the SNR, decreasing from $19.80$ to $8.86$ with the presence of thermal noise. 
The filament detection results based on various galaxy-masking approaches are presented in \reftb{tab: diff mask result compare}.

Compared to the "Unmasked" case, the "Mask MGS" scenario exhibits only an approximate $2\,\%$ decrease in signal, whereas the "Mask All" approach results in about a $60\,\%$ drop in signal. 
This significant amplitude reduction observed with the aggressive galaxy masking method indicates that faint galaxies, particularly those residing in or near the filament, contribute substantially to the \hi brightness temperature. 
When comparing to the filament \hi only results, we find that the contributions from galaxies still account for a considerable portion of the total signal, even in the "Mask All" case.
This highlights the critical importance of selecting an appropriate frequency width during the masking process.

\subsection{Impact of frequency width}\label{sec: multi_slices}
Considering the spectral broadening caused by the peculiar velocity, varying the frequency width choice would affect our filament signal estimation.

\subsubsection{The impact on pairwise stacking}

\begin{figure*}
    \centering
    \includegraphics[width=\textwidth]{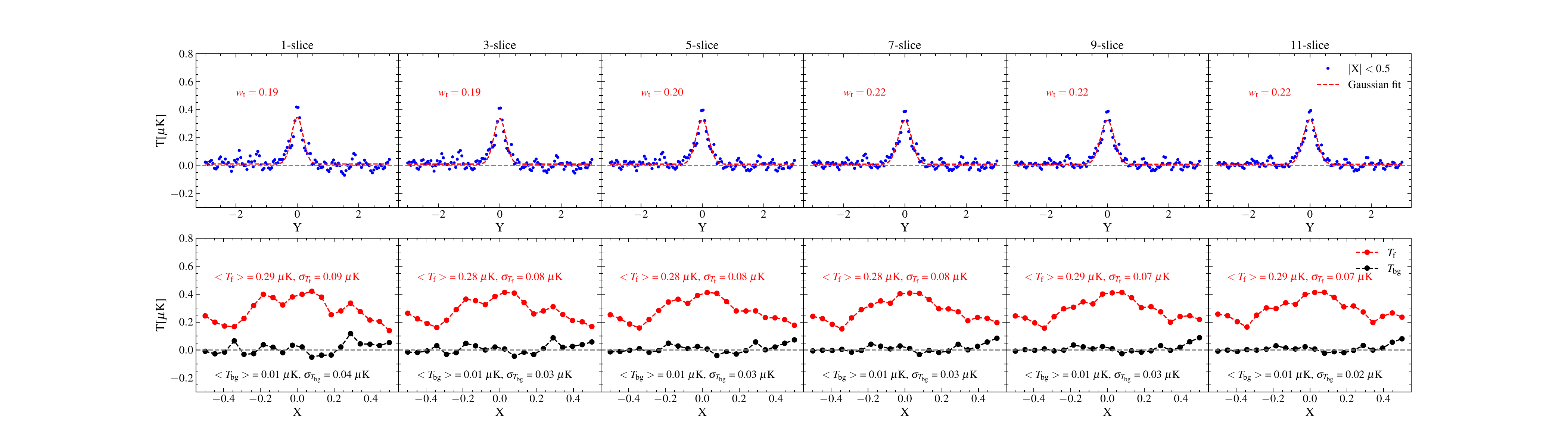}
    \caption{The comparison of filament profiles conducted by varying average numbers of frequency slices stacked on the filament \hi only map. The top and bottom panels display the transverse and longitudinal profiles, respectively, similar to those shown in \reffg{fig: fuzz_profiles}.}
    \label{fig: fuzz_stack_compare}
\end{figure*}

\begin{figure}
    \centering
    \includegraphics[width=0.5\textwidth]{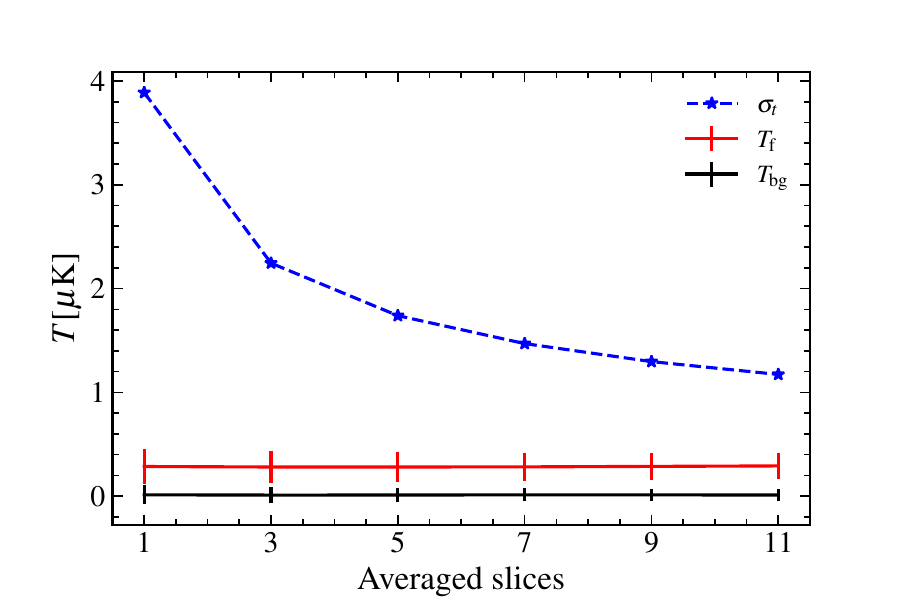}
    \caption{Comparison of the mean signal levels across different averaged numbers of frequency slices stacked on the filament \hi only map. Error bars represent the mean and standard deviation of filament (red) and background (black) values. 
    The dashed blue line indicates the standard deviation of the thermal noise, obtained in a $48\,{\rm s}$ integration time, $\sim200\,\deg^2$ sky area of FAST \hiim survey.}
    \label{fig: fuzz_level_multi}
\end{figure}

In addition to the central 1-slice case previously mentioned, we also estimate the filament signals for stacks consisting of 3 slices, 5 slices, 7 slices, 9 slices, and 11 slices, respectively.

As shown in \reffg{fig: fuzz_stack_compare}, with the filament \hi only map, 
the estimated filament width and the average signal amplitude increase marginally with wider frequency widths. 
Conversely, the background noise level exhibits a decreasing trend as the number of slices increases.
These findings suggest that the filament signal is uniformly distributed across the 11 frequency slices (spanning $1.1\, \mhz$). 

To investigate the impact of instrumental thermal noise, we attempted to extract the filament signal from the
filament only map with noise injected. 
However, in our $48\,{\rm s}$ $200\,\deg^2$ FAST \hiim survey simulation, thermal noise overwhelmingly dominated 
the filament signal, rendering it undetectable.
Instead, we calculated the standard deviation of the stacked noise-map result to quantify the thermal noise level, 
depicted by the blue dashed line in \reffg{fig: fuzz_level_multi}. 
For comparison, the filament and background amplitudes as the function of averaged frequency slices are shown 
in red and black, where the error bars indicate the standard deviation, respectively.
As illustrated, the thermal noise level exceeds the filament signal by an order of magnitude in the 1-slice case. 
Although thermal noise decreases with increasing frequency width, it remains approximately \add{fourfold} as large as 
the filament signal even in the 11-slice case.
This indicates that additional integration time is required to reduce the thermal noise level comparable to the filament signal.

\subsubsection{The impact on masking galaxy contribution}

\begin{figure*}
    \centering
    \includegraphics[width=\textwidth]{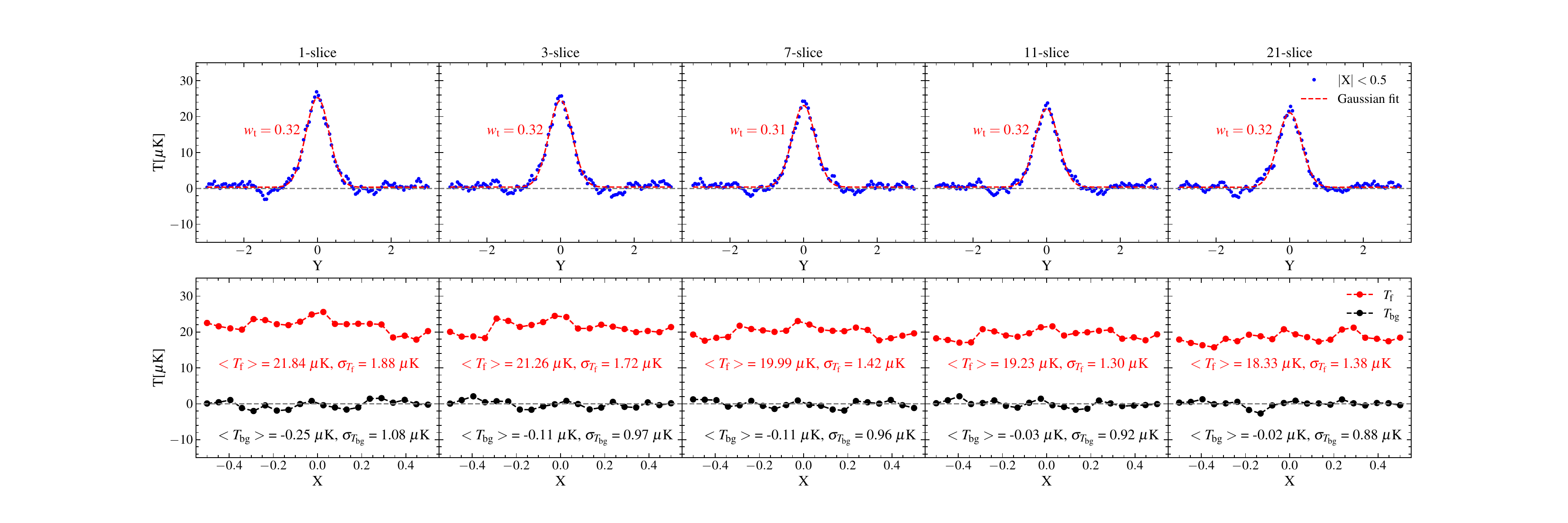}
    \includegraphics[width=\textwidth]{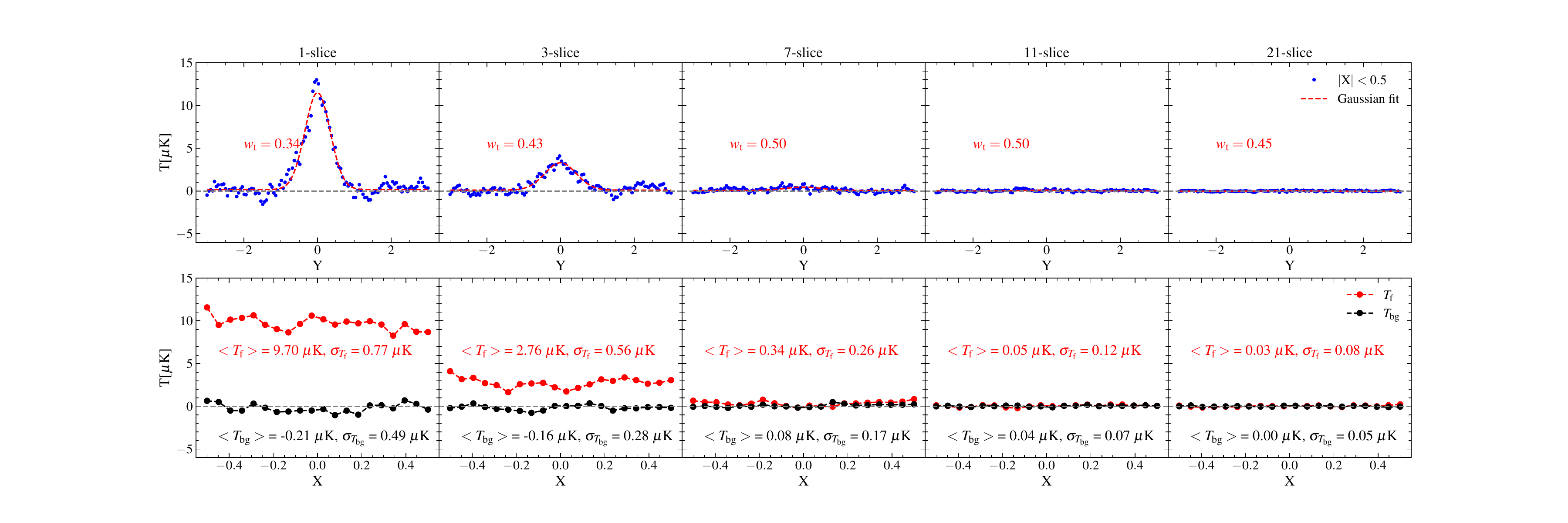}
    \caption{The comparison of different numbers of masking slices. The top two and bottom two panels show the "Mask MGS" and "Mask all" cases, respectively. \textit{Top}: The transversal-section profile (solid blue) and its best-fitted Gaussian curve (dashed red). 
    \textit{Bottom}: the lengthwise-section profiles of filament (red) and background (black).
    }
    \label{fig: mask_compare}
\end{figure*}
    
As demonstrated in Section \ref{sec: galaxy_contributed}, galaxy contributions predominantly influence filament reconstruction, highlighting the crucial role of masking slice selection.
We now compare the impact of varying frequency slice numbers on filament reconstruction for 
two masking scenarios: "Mask MGS" and "Mask All".

As illustrated in \reffg{fig: mask_compare}, the "Mask MGS" case exhibits a moderate decline in signal amplitude and standard deviation with increasing frequency width. 
Notably, even in the 21-slice case, these values remain significantly higher than the previously estimated average filament signal.
In contrast, the "Mask All" scenario reveals a steep downward trend, with signal amplitude and standard deviation dropping from $9.70\, \mu{\rm K}$ (1-slice) to $2.76\, \mu{\rm K}$ (3-slice), ultimately becoming negligible. 
This reinforces the notion that faint galaxies, rather than bright MGS-like galaxies, predominantly influence filament signal estimates.
It also emphasizes the importance of selecting an appropriate masking frequency width to reduce the contribution of these galaxies.
However, achieving this requires a deep galaxy survey that covers the \hiim region of interest, providing more detailed information about the faint galaxies, including their positions and frequency widths. 
This goal may be attainable with powerful instruments like the SKA in the future.

Our current pairwise stacking analysis relies on an optically selected galaxy catalog, 
which is well-suited for tracing galaxy cluster centers over a wide range of redshifts. 
However, galaxies selected in the radio band are expected to form a distinct sample compared to the optically selected ones. 
With upcoming large-area deep \hi surveys, an \hi emission line-selected galaxy sample could offer 
an alternative set of tracers for such analyses. We will explore these possibilities in future work.

\subsection{Filament extension}

\begin{figure*}
    \centering
    \includegraphics[width=\textwidth]{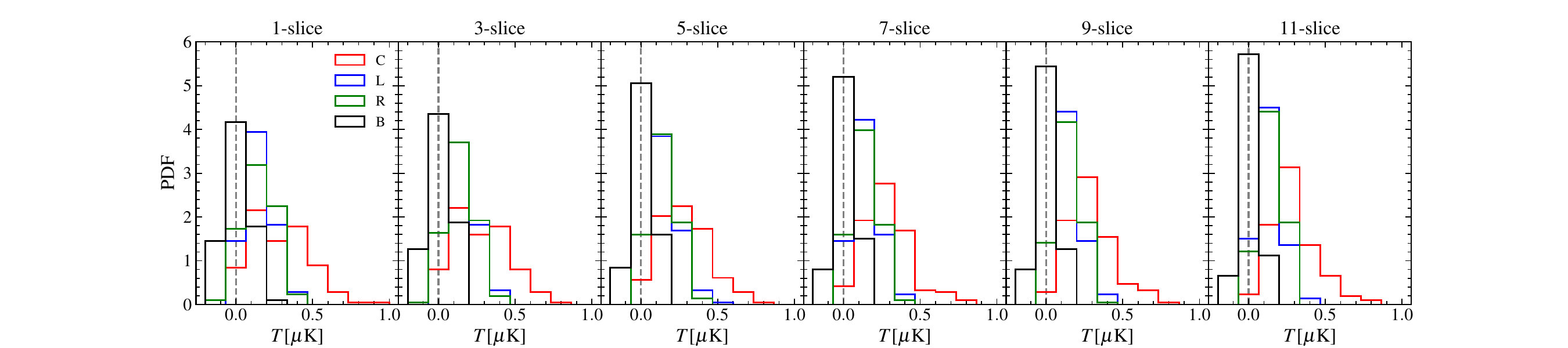}
    \caption{
    The histogram comparison of varying averaged numbers of frequency slices in stacking on the filament \hi only map. 
    Each histogram represents the distributions of the brightness temperature within the four selected regions in the 2D-PSM, as defined in \refeq{eq:area}.
    The gray dashed line in each panel markers $T=0$.
    }
    \label{fig: extension_fuzz_compare}
\end{figure*}

\begin{figure}
    \centering
    \includegraphics[width=0.5\textwidth]{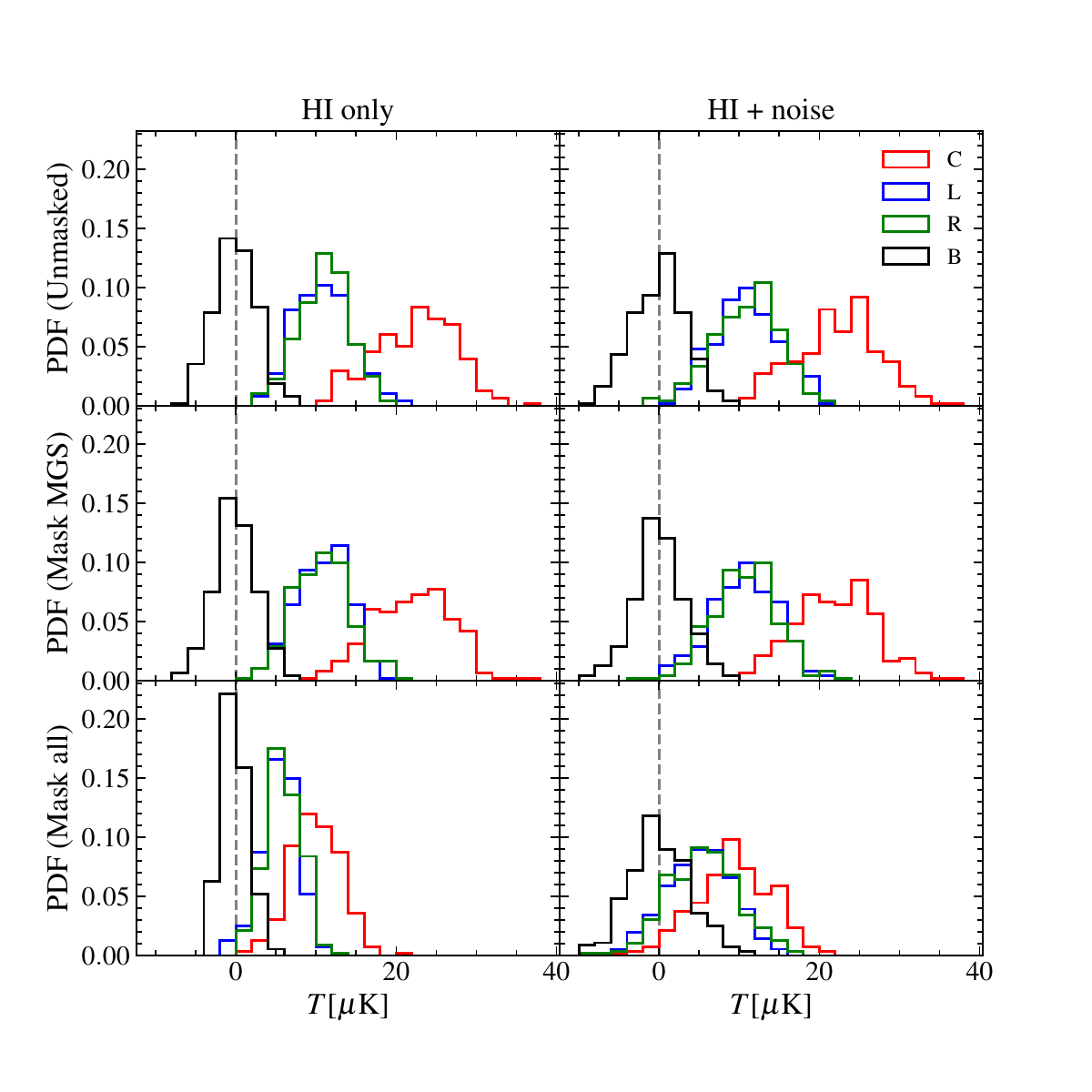}
    \caption{ 
    A histogram comparison similar to \reffg{fig: extension_fuzz_compare}, but illustrating three different galaxy contribution masking approaches. The rows represent the "Unmasked" (top), "Mask MGS" (middle), and "Mask All" (bottom) cases, respectively. The results without observational noise are shown in the left panels, while the right panels display the results with noise.
}
    \label{fig: extension_mask_compare}
\end{figure}

The filament signal in the final 2D-PSM shows a clear extension, which indicates that
\hi does not only reside between the galaxy pairs but also extends outside the pair along its axis.
To quantify the brightness distribution, we then select four regions in the 2D-PSM, according to the following scheme:
\begin{eqnarray}
\begin{aligned}\label{eq:area}
{\rm C} :& &       |&X| < 0.5 ,              &|Y|& < w_{\rm t}, \\
{\rm L} :& &-1.5 < &X < -0.5,              &|Y|& < w_{\rm t},  \\
{\rm R} :& &0.5 <  &X < 1.5 ,              &|Y|& < w_{\rm t},\\
{\rm B} :& &       |&X| < 0.5 ,  &3w_{\rm t} <|Y|& < 4w_{\rm t},
\end{aligned}
\end{eqnarray}
where C represents the region residing inside the galaxy pair and B represents the background region. 
These are the same as those we employed to quantify the lengthwise section filament and the background 
profiles in \refsc{sec:quantification}.
L and R, instead, represent the areas residing to the left and right of the pair, 
along the extension of the pair axis. 

The brightness temperature histograms for the four selected regions are shown in \reffg{fig: extension_fuzz_compare} and \reffg{fig: extension_mask_compare}.
As we show in \reffg{fig: extension_fuzz_compare}, there is significant \hi existence in the area between the galaxy pairs and along their extension lines.
While slightly lower in amplitude, the signal level of the extension part exhibits a similar trend to that of the central filament signal, characterized by a stable mean value and narrower distribution across larger cases of stacked slice configurations.
We also presented results illustrating the contributions from galaxies, as shown in \reffg{fig: extension_mask_compare}. 
The comparison of results without masks, with MGS-like galaxy masks applied, and with full galaxy sample masked are displayed in the top, middle, and bottom panels respectively.
As more galaxies contribute to the signal, the level of the extension part increases.
The mean level of the extension is approximately half that of the filament across all cases.

\add{Our selection criteria require galaxy pairs to be located in different clusters. 
However, these clusters are typically positioned at cosmic nodes, which are interconnected by multiple filaments. 
As a result, while each galaxy pair is primarily connected by a central filament, i.e. the area C in its 2D-IPM, 
additional filaments extending from the same nodes contribute to the observed extensions. 
The candy shape may suggest that filaments emerging from a common node tend to align along the same line.
}

\subsection{Background level}\label{sec: bglevel}

\begin{figure}
    \centering
    \includegraphics[width=0.48\textwidth]{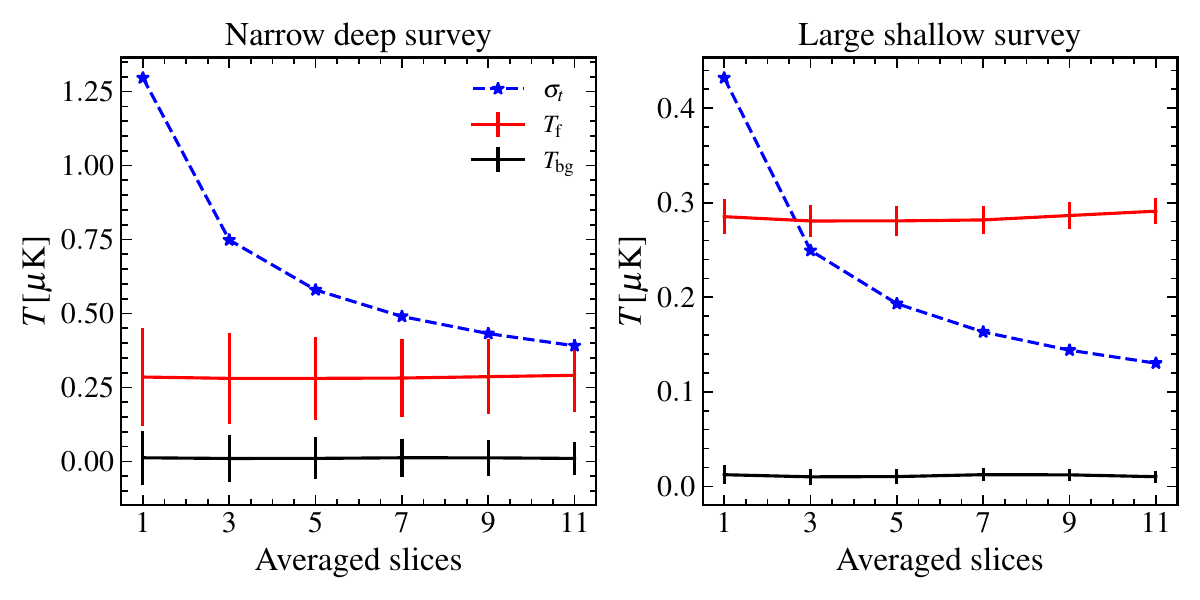}
    \caption{Comparison of the mean signal levels similar to \reffg{fig: fuzz_level_multi}, but showing results from a narrow-field deep survey (left, \add{$\sim 200\,{\deg}^2$, $432\,{\rm s}$ per beam}) and a large-field shallow survey (right, \add{$\sim 1800\,{\deg}^2$, $48\,{\rm s}$ per beam}) (see Section \ref{sec: bglevel} for details).
    }
    \label{fig: fuzz_level_future_multi}
\end{figure}

In addition to instrumental thermal noise, random neutral hydrogen signals from nearby galaxies contribute to the background level.
The contributions from various types of galaxies differ significantly according to their size, 
velocity, and distance from the filament. 
This variability, which we refer to as background variation, plays a crucial role in extracting the \hi signal.
After masking galaxy contributions within a fixed radius of $\sim 120\,{h}^{-1}{\rm kpc}$, 
the residual background primarily comes from the effect of galaxy spectrum broadening. 
This spectrum broadening introduces the galaxy contribution even for galaxies far from the filaments.

Notably, different galaxy-masking approaches result in varying background levels. 
As demonstrated in \reftb{tab: diff mask result compare}, the background level does not decrease when the contribution from MGS-like galaxies (luminous galaxies detectable by current optical surveys) is removed. 
This is because MGS-like galaxies, while luminous, are fewer in number and do not contribute significantly to the overall background.
In contrast, numerous faint galaxies, with a more random positional distribution and a broader \hi mass distribution, exert a much greater influence on background variations. 
Their contribution to the background is more substantial and harder to mitigate than that of MGS-like galaxies.

The background level can be reduced with substantial integration time of the pixels in the final 2D-PSM.
Given the same total observation time, it can be achieved by a small-field deep survey, which is to
continue scanning the same area of the sky.
The small-field deep survey can significantly reduce the thermal noise-induced background variation.
However, it can not reduce the random galaxy-induced background variation because 
of the fixed survey area.
Instead, the \hi survey can be carried out with a large-area shallow survey. 
With the stacking analysis, both the thermal-induced and random galaxy-induced background can be reduced simultaneously. 

This work assumes a survey area of $\sim 200\,{\deg}^2$ and an integration time of $\delta t=48\,{\rm s}$ per beam. 
The $\sim 200\,{\deg}^2$ sky area allowed us to select $3526$ MGS-like galaxies.
These galaxies form \mgspairs galaxy pairs meeting our separation criteria, with a pairing rate of approximately $1.23\,\%$. 
Assuming the galaxy distribution is approximately uniform on large scales, we estimate that 
doubling the survey area would result in a fourfold increase in the size of the galaxy pair catalog. 
This scaling shows that there would be a significant reduction in the background level if we increased the 
number of pairs used in the stacking process.
Using an optimized survey area at $1800\,\deg^2$, which overlaps between the SDSS and FAST fields of view, i.e. within a Zenith angle of $15^\circ$ at the FAST site \citep{2020RAA....20...64J}, the 
background level is expected to be reduced by about one order of magnitude.

In \reffg{fig: fuzz_level_future_multi}, we compare the thermal noise levels for future large shallow surveys (right panel) 
and narrow deep surveys (left panel) with the filament signal levels and background variations derived from a 
filament \hi only map.
In the case of a narrow deep survey, the thermal noise and filament signal levels are comparable. 
This suggests that, despite reducing the thermal noise, the deep survey does not provide a significant 
advantage in terms of SNR for detecting filaments, as the noise remains at the same level as the signal.
In contrast, the thermal noise for a large shallow survey is significantly lower than the filament signal level. 
This indicates that the large shallow survey strategy, which covers a larger sky area, is more effective for detecting filaments through the pairwise stacking method.

\subsection{Filament \hi column density}

According to the statistical analysis of the physical distance between galaxy pairs in the MGS-like catalog,
the mean transversal separation is $\sim 10.1\ h^{-1}{\rm Mpc}$, which corresponds to the distance between 
$(1,0)$ and $(-1,0)$ in the final 2D-PSM. Thus, the unit of the 2D-PSM is scaled on average by $10.1/2\ h^{-1}{\rm Mpc}$.

The filament radius is estimated as,
\begin{equation}\label{eq:filamentwidth}
    r = w_{\rm t} \times \frac{10.1\ h^{-1}{\rm Mpc}}{2} \approx w_{\rm t} \times 5.05\ h^{-1}{\rm Mpc},
\end{equation}
where $w_{\rm t}$ is the width of the Gaussian profile as defined in \refeq{eq:profile}.

\add{As shown in \reftb{tab: diff mask result compare}, the filament radii are increased, from $\sim 0.96\ h^{-1}{\rm Mpc}$ in the filament only case, to $\sim 1.61\ h^{-1}{\rm Mpc}$ in the cases with the contribution of galaxies.
The estimation of the filament radius is based on the mean value of the filaments
selected by the galaxy pairs. As shown in \reffg{fig: MGSpairs_sep}, the separation distances of the 
galaxy pairs are uniformly distributed with a mean value of $\sim 10.1\ h^{-1}{\rm Mpc}$
and standard deviation of $\sim 2.5\ h^{-1}{\rm Mpc}$.
Such scattering of the separation distances could be propagated to the final estimation of the
filament radius,
\begin{align}
\sigma_r = w_{\rm t} \times \frac{2.5\ h^{-1}{\rm Mpc}}{2} \approx w_{\rm t} \times 1.25\ h^{-1}{\rm Mpc},
\end{align}
and results in an uncertainty of $0.24\, h^{-1}{\rm Mpc}$ for the filament-only case
and $0.4\, h^{-1}{\rm Mpc}$ with galaxy contribution, respectively.
}

\add{We note that our estimation of the filament radius assumes a constant radius, 
independent of the filament’s length. This is a simplifying assumption that may 
introduce uncertainties. Future analyses should account for potential variations 
in filament radius with length, as longer filaments tend to be thicker 
(see, e.g., \cite{2024MNRAS.532.4604W}).}

For a single, homogeneous, optically thin \hi cloud, the column density $N_{\hi}$ is related 
to its velocity-broadened profile width $\Delta v$ and the filament brightness temperature 
$T_{\rm f}$ via:
\begin{equation}\label{eq: NHI_dv}
    \Bigg(\frac{N_{\hi}}{{\rm cm^{-2}}}\Bigg) = 1.82\times10^{12}\
    \Bigg(\frac{T_{\rm f}}{\rm \mu K}\Bigg) 
    \Bigg(\frac{\Delta v}{\rm km\ s^{-1}}\Bigg).
\end{equation}

In our 1-slice case, the corresponding width is about $\Delta v = 20\ {\rm km\ s}^{-1}$.
Substituting the results of noisy "Unmasked" galaxy-contributed case, i.e. with the values of $T_{\rm f}=22.26\pm2.11\ \mu{\rm K}$,
we obtain an \hi column density of $N_{\hi}=8.10\pm0.77\times10^{14}\ {\rm cm^{-2}}$.
For the filament only case, we substitute the $T_{\rm f}=0.29\pm0.09\ \mu{\rm K}$ into \refeq{eq: NHI_dv}, and obtaining an \hi column density of $N_{\hi}=1.06\pm0.33\times10^{13}\ {\rm cm^{-2}}$, which is about two orders of magnitude lower than that of the case with galaxy contribution.

We can also estimate the filament \hi density parameter, defined as the ratio between the filament \hi density $\rho_{\hi}(z)$ and the critical mass density $\rho_{\rm c}(0)$:
\begin{equation}\label{eq_density_parameter}
\begin{split}
    \Omega^{\rm f}_{\hi}(z) = &\frac{\rho_{\hi}(z)}{\rho_{\rm c}(0)} \\
    = &7.6\times 10^{-3}\ \Bigg(\frac{T_{\rm f}}{\rm mK}\Bigg) \Bigg(\frac{h}{0.7}\Bigg)^{-1} (1+z)^{-2} E(z),
\end{split}
\end{equation}
where $E(z)=[\Omega_{\rm m}(1+z)^{3} + \Omega_{\Lambda}]^{1/2}$.
Substituting the noisy "Unmasked" galaxy-contributed case again results in an estimation of the filament \hi density parameter as $\Omega^{\rm f}_{\hi}=1.52\pm0.14\times 10^{-4}$ at $z = 0.1$. 
As a comparison, the filament only case has an \hi density parameter of $\Omega^{\rm f}_{\hi} =1.98\pm0.61\times 10^{-6}$ at $z = 0.1$, which is again lower by two orders of magnitude.

\section{Conclusions}
\label{sec:conclusion}

This work presents an end-to-end simulation aimed at identifying \hi in filament structures 
utilizing FAST \hiim survey with a galaxy pairwise stacking analysis. 
The simulation started by constructing an \hi brightness temperature sky cube for both the filament only and with the galaxy contribution, and the
corresponding galaxy optical survey catalog. 
The SDSS DR7 Main Galaxy Sample (MGS)
selection criteria are applied to simulate an MGS-like galaxy catalog, which has both the redshift distribution and sky footprint overlapping with the FAST \hiim drift scan survey
\citep{2020MNRAS.493.5854H,2023ApJ...954..139L}. 
This work assumed an observational sky area of $\sim 200\,{\deg}^2$ and an integration time of $48\,{\rm s}$ per beam.

We showed that, in the filament only case, after subtracting the \hi brightness temperature contribution from the halos, a significant filament structure is recovered in the final 2D pairwise-stacked map (2D-PSM), which indicates a substantial presence of \hi gas residing in the filament connecting galaxy pairs.
It yields an average \hi filament signal amplitude of $\sim 0.29\ {\mu{\rm K}}$ at $z\simeq 0.1$ with an SNR of $3$, indicating the ability of filament reconstruction through pairwise stacking methods.
We also showed that the estimated filament is stable across different choices of frequency slices with $1.1\,\mhz$, which enables tailored optimization for SNR improvement.

With the contribution of those nearby galaxies, however, the mean brightness temperature is increased to $\sim 22.4\ {\mu{\rm K}}$, about two orders of magnitude larger.
We further quantified the fraction of the \hi contribution 
from these extra galaxies by applying different levels of galaxy masking. 
Masking the MGS-like galaxies results in 
about $2\%$ drop in signal, while masking all potentially existing galaxies results in a signal drop of about $60\%$, indicating the signal is dominantly contributed by faint sources.
By increasing the number of masking slices, we reaffirm the dominant contribution of faint galaxies to the signal and highlight the critical importance of selecting an appropriate frequency width to effectively mask their contributions. However, this approach requires a deep galaxy survey covering the \hiim region of interest, which may be made possible with future advanced instruments like the SKA.

In the filament only case, current pilot observations for the FAST \hiim drift scan survey are dominated by thermal noise.
In addition to the observational thermal noise, measurement uncertainty is also caused by the background variation, which is due to the brightness leakage from the additional galaxies
randomly located around the filaments.
Given a fixed total observation time, a wide-field \hiim survey, which includes a large number of galaxy pairs, can simultaneously reduce thermal noise to below the filament signal level and minimize background variation to a negligible level. 

Using the filament brightness temperature in the final 2D-PSM, we estimated the filament \hi column density to be $N_{\hi}=1.06\pm0.33\times10^{13}\ {\rm cm^{-2}}$ at a redshift of $0.1$ for the filament only case. However, when the contribution of galaxies is included, the column density increases to $N_{\hi}=8.10\pm0.77\times10^{14}\ {\rm cm^{-2}}$.
We also estimated the \hi mass density of the filament. 
In the filament only case, the mass density is $\Omega^{\rm f}_{\hi} =1.98\pm0.61\times 10^{-6}$. However, when the contribution of galaxies is included, this value increases to $\Omega^{\rm f}_{\hi}=1.52\pm0.14\times 10^{-4}$.
These results indicate that the contribution of galaxies leads to an overestimation of the \hi column density and mass density in filaments by approximately two orders of magnitude.

\section*{Acknowledgements}
We acknowledge the support of the National SKA Program of China (Nos.~2022SKA0110200, 2022SKA0110203, 2022SKA0110100, 2022SKA0110101),
the NSFC International (Regional) Cooperation and Exchange Project (No. 12361141814),
and the National Natural Science Foundation of China (Nos. 12473091, 11975072, 11835009).
DT acknowledges financial support from the XJTLU Research Development Fund (RDF) 
grant with number RDF-22-02-068.

\section*{Data Availability}

The data underlying this article will be shared on reasonable request to the corresponding author.

\bibliography{ref}
\bibliographystyle{aasjournal}

\appendix
\section{Uncertainty in the halo profile subtraction}\label{sec: halo_sub}

\add{
The \hi halo contribution shown in the 2D-PSM (see the middle panel of \reffg{fig: illu_results} for example) is actually a superposition of re-scaled halos, making it vastly different from any realistic halo profile.
Instead of deriving an analytical halo profile, we use the numerical method, presented in
\refsc{sec:sub_halo}, to obtain the best-fit profile of the halo contribution.
}

\subsection{Jackknife error for halo profile estimation}

\begin{figure}
    \centering
    \begin{minipage}[t]{0.48\textwidth}
        \vspace{0pt}  
        \centering
        \includegraphics[width=\textwidth]{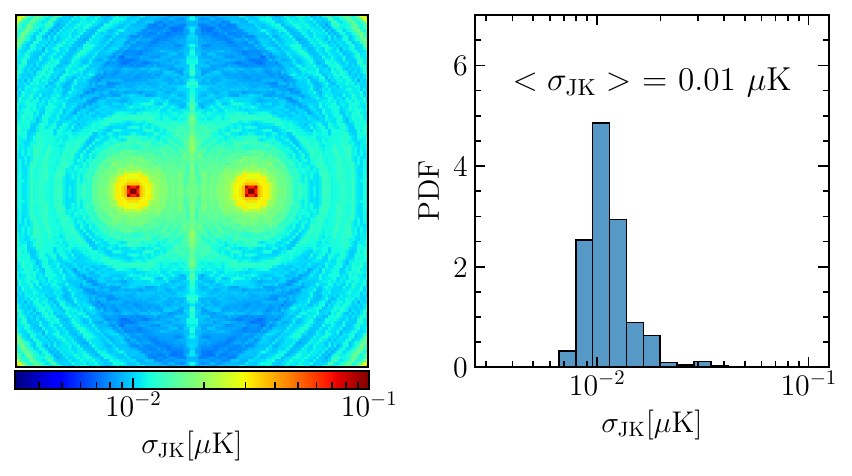}
        \caption{The jackknife estimation of the standard error in halo fitting for the case of filament \hi only. The standard error and its probability density distribution are shown in the left and right panels, respectively. The mean value of the standard error is labeled in the right panel and serves as the reference for the uncertainty introduced by halo fitting.}
        \label{fig: halo_error}
    \end{minipage}
    \hfill
    \begin{minipage}[t]{0.48\textwidth}
        \vspace{0pt}
        \centering
        \includegraphics[width=\textwidth]{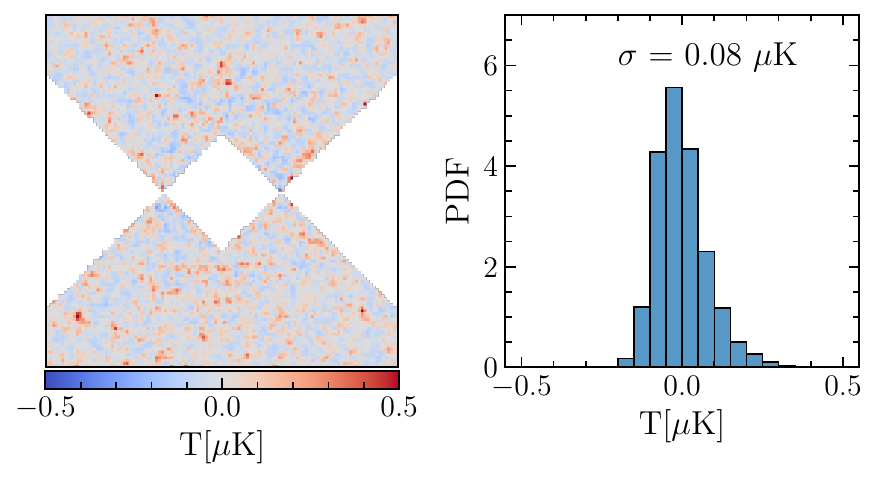}
        \caption{
        The residuals of halo subtraction for the case of filament \hi only. The halo residuals within the fitted area and their probability density distribution are shown in the left and right panels, respectively.
        }
        \label{fig: halo_res_f}
    \end{minipage}
\end{figure}

\add{
To assess the accuracy of the halo profile estimation, we use jackknife samples
(see e.g. \cite{2016arXiv160600497M}) to evaluate the estimation error.
As described in \refsc{sec:method}, the full galaxy pair sample is divided into 
$153$ sub-samples and construct the 2D-PSM, individually. 
These $153$ 2D-PSMs are then averaged to form the final 2D-PSM, which is used for
the rest analysis, including the halo profile estimation.
The error in the halo profile estimation is evaluated as,
\begin{align}\label{eq:jackknife_error} 
\sigma_{\rm JK} = \sqrt{\frac{n-1}{n}\sum_{i=1}^{n}[\hat{\theta}_{(i)}-\hat{\theta}_{(.)}]^2},
\end{align}
where $n=153$ is the number of jackknife samples, 
$\hat{\theta}_{(i)}$ is the halo profile estimated without the $i$-th sample, and
$\hat{\theta}_{(.)}$ is the empirical average of these halo profile estimations,
\begin{align}\label{eq:jackknife_rep} 
\hat{\theta}_{(.)} = \frac{1}{n}\sum_{i=1}^{n}\hat{\theta}_{(i)}.
\end{align}
}

\add{
The estimation error evaluated using the filament-only \hi sky map is shown 
in \reffg{fig: halo_error}, 
where the left panel shows the errors of each 2D-PSM pixel and the right panel shows the histogram 
distribution of pixel errors. 
The errors are significantly amplified near the halo center and the edges, 
since few pixels are included for halo-profile estimation. 
We use the mean of the errors, i.e. $\langle\sigma_{\rm JK}\rangle = 0.01\,{\rm \mu K}$,
as the reference for the estimation uncertainty.}

\add{The jackknife error represents the variance raised from the scattering of individual 
halo profile and it is negligible compared to the absolute filament brightness temperature 
$T_{\rm f} = 0.29 \pm 0.09 \, {\rm \mu K}$ (as listed in \reftb{tab: diff mask result compare}).
}

\subsection{Variance for halo profile subtraction residuals}
\add{
As a comparison, we extract the pixel residuals, i.e. the 2D-PSM after the halo profile subtracted, 
within the halo-profile-fitting area and adopt the residual errors as the 
intrinsic uncertainty $\sigma_{\rm h}$. 
The result for the filament-only case is shown in \reffg{fig: halo_res_f}. 
We observe a zero-mean Gaussian distribution with a standard deviation 
$\sigma_{\rm h} =0.08\, \mu{\rm K}$.
}

\add{
The residual variance $\sigma_{\rm h}$ arises from the random \hi emissions
within or close to the halo. In \refsc{sec: bglevel}, we selected an relatively empty 
area to evaluate the background \hi variance. In the case of using the filament-only map,
the background variance is about 
$\sqrt{n_{\rm p}} \sigma_{T_{\rm bg}} = 0.11\,{\rm \mu K}$, where $n_{\rm p}$ is the 
number of pixels averaged along the $Y$-axis in the B area (as defined in \refeq{eq:area}).
The residual variance $\sigma_{\rm h}$ is close to the background variance, which 
indicates a significant elimination of halo contribution. 
} 

\subsection{The galaxy contribution and thermal noise}

\begin{table}
    \caption{
    Comparison of the jackknife errors and the residual errors.
    $\langle\sigma_{\rm JK}\rangle$ represents the mean jackknife error of the halo profile
    estimation, 
    while $\sigma_{\rm h}$ denotes the standard deviation of the halo residual within the 
    halo-profile-fitting sector. 
    $T_{\rm f}$ corresponds to the filament \hi brightness temperature, 
    as quoted from \reftb{tab: diff mask result compare}. 
    The term $\sqrt{n_{\rm p}}\sigma_{T_{\rm bg}}$ represents the standard deviation 
    of background brightness temperature fluctuations, 
    where $n_{\rm p}$ is the number of pixels within the filament width.}\label{tab: compare_halo}
    \centering
    {\scriptsize
    \begin{tabular}{clccccccc}
    \hline
    && filament \hi only & Unmasked & Mask MGS & Mask all & Unmasked (noise) & Mask MGS (noise) & Mask all (noise)\\
    \hline
    $\sigma_{\rm JK}$ & [$\mu$K] & 0.01 & 0.40 & 0.40 & 0.27 & 0.47 & 0.47 & 0.47 \\
    $\sigma_{\rm h}$  & [$\mu$K] & 0.08 & 5.18 & 3.30 & 1.95 & 5.65 & 3.99 & 3.87 \\
    $T_{\rm f}$ & [$\mu$K] & $0.29\pm0.09$ & $22.20\pm2.17$ & $21.84\pm1.88$ & $ 9.70\pm0.77$ & $22.26\pm2.11$ & $21.92\pm1.87$ & $ 9.13\pm1.22$ \\
    $\sqrt{n_{\rm p}}\sigma_{T_{\rm bg}}$ & [$\mu$K] &0.11 & 3.95 & 3.74 & 1.70 & 3.89 & 4.02 & 3.57 \\
    $n_{\rm p}$ & & 8 & 12 & 12 & 12 & 12 & 12 & 12 \\
    \hline
    \end{tabular}
    }
\end{table}

\add{In the cases with galaxy \hi contribution, as well as thermal noise, both the
jackknife errors and the residual errors are amplified. We list all the results in 
\reftb{tab: compare_halo}.
Generally, the jackknife errors in all cases are negligible compared to the absolute 
\hi brightness temperature of the filament. The jackknife errors are also smaller than
the residual variance, which represents the intrinsic errors induced by the background variance. 
Although the halo profile estimation induces uncertainties, it provides significant 
accuracy, giving the current intrinsic errors. 
}

\end{document}